# Caulking the "Leakage Effect" in MEEG Source Connectivity Analysis

Eduardo Gonzalez-Moreira[1,3,4#], Deirel Paz-Linares[1,2#], Ariosky Areces-Gonzalez[1,5], Rigel Wang[1], Jorge Bosch-Bayard[4], Maria Luisa Bringas-Vega[1,2] and Pedro A. Valdés-Sosa[1,2*]

**#** *contributed equally as first authors*

deirel.paz@neuroinformatics-collaboratory.org

eduardo.g.m@neuroinformatics-collaboratory.org

**\*** *corresponding author*

pedro.valdes@neuroinformatics-collaboratory.org

Affiliations:

(1) The Clinical Hospital of Chengdu Brain Science Institute, MOE Key Lab for Neuroinformation, University of Electronic Science and Technology of China, Chengdu, China
(2) Cuban Neuroscience Center, La Habana, Cuba
(3) Centro de Investigaciones de la Informática, Universidad Central "Marta Abreu" de Las Villas
(4) Departamento de Neurobiología Conductual y Cognitiva, Instituto de Neurobiología, Universidad Nacional Autónoma de México. Boulevard Juriquilla 3001, Querétaro, 76230, México
(5) Departamento de Informatica, Universidad de Pinar del Rio, Pinar del Rio, Cuba.

## Summary

Simplistic estimation of neural connectivity in MEEG sensor space is impossible due to head volume distortion. The only viable alternative is to carry out connectivity estimation in source space. Among the neuroscience community this is claimed to be impossible or misleading due to "Leakage": linear mixing of the reconstructed sources. To address this problematic we propose a novel solution method that "caulks" the "Leakage" in MEEG source activity and connectivity estimates: Brain Connectivity Variable Resolution Tomographic Analysis (BC-VARETA). It is based on a joint estimation of source activity and connectivity in the frequency domain representation of MEEG time series. To achieve this, we go beyond current methods that assume a fixed gaussian graphical model for source connectivity. In contrast we estimate this graphical model in a Bayesian framework by placing priors on it, which allows for highly optimized computations of the connectivity, via a new procedure based on the local quadratic approximation under quite general prior models. A further contribution of this paper is the rigorous definition of leakage via the Spatial Dispersion Measure and Earth Movers Distance, based on the geodesic distances over the cortical manifold. Both measures are extended for the first time to quantify "Connectivity Leakage" by defining them on the cartesian product of cortical manifolds. Using these measures, we show that BC-VARETA outperforms most state of the art inverse solvers by several orders of magnitude.

## Highlights

- **Nonlinear method BC-VARETA allows to caulk the leakage in MEEG source activity and connectivity estimates.**
- **Optimized solution of gaussian graphical models with general penalization function via local quadratic approximation.**
- **Quantification of the connectivity leakage by the extension of spatial dispersion measure and earth movers distance to cartesian product of cortical manifold spaces.**
- **Demonstrated the inefficacy of MEEG sensors space connectivity analysis by comparison against source space analysis with BC-VARETA**

# 1 Introduction

The estimation of neural connectivity from EEG or MEG data is at the crossroad today. The essential debate is whether these estimates should be obtained in sensor space or source space and the limitations of each of these approaches, see a discussion on this topic in (Palva, et al., 2018). Equating neural connectivities in brain networks to the statistical dependencies at the sensors space is a common fallacy (Blinowska, 2011). It is rendered invalid by two combined effects: head volume distortive inhomogeneities and mixture of source activity from the whole brain (Brunner et al., 2016). Alternatively, there has been a quest for measures that somehow 'ameliorate' the volume distortive effect (Kaminski and Blinowska, 2014; Kaminski and Blinowska, 2017). This is also an unlikely enterprise. None of these procedures use explicitly knowledge about the Forward Model or Lead Field to cancel its effect (Van de Steen et al., 2016).

It would thus seem that the only sensible procedure to estimate neural connectivity would be to analyze interactions between estimated sources given the Forward Model inversion (Inverse Problem or Electrophysiological Sources Imaging). While attempting to counteract the volume distortive effect, with Electrophysiological Sources Imaging methods, also suffers from two difficulties. First, for different reasons, there are neural generators whose activity is not reflected at the sensors. Second, any of the methods to estimate sources suffers from the "Leakage Effect".

The first problem refers to that invisible sources can only be encountered by prior knowledge encoded into the source activity and connectivity estimation procedure (Krishnaswamy et al., 2017). The second, Leakage (Schoffelen and Gross, 2009), refers to a blurred reconstruction of point sources that entails spillover of activity between them thus distorting the estimates of their inter-connections. Leakage is a well-known problem in all medical imaging techniques but is much more severe for MEEG source reconstruction methods. It is not surprising that there are many attempts to modify MEEG inverse methods to ameliorate or avoid source Leakage (Freeman, 1980; Brookes et al., 2012; McCoy and Troop, 2013; Colclough et al., 2015; Wens et al., 2015; Colclough et al., 2016; Silva Pereira et al., 2017; Hedrich et al., 2017, Farahibozorg et al., 2018).

Unfortunately, a difficulty in evaluating "Leakage Correctors" is the lack of a direct metric of the distortions in connectivity. Rather, what exists are measures of Leakage distortion of source estimators--not connectivity. One such measure is the dispersion of the "Point Spread Function" (PSF)—the reconstruction of a point sources. There is no doubt that reducing the distortions in activation, i.e. Type I Leakage, will be a good thing for connectivity estimates, but much better would be a direct measure of the distortion in connectivity, i.e. Type II Leakage. Another difficulty towards Leakage correction, is that the most stablished methods are based on connectivity postprocessing of estimated source activity given by a source localization procedure. Thus, they do not make use of more consistent models of sources activity and connectivity, i.e. system identification. In this sense the state of the art of sophisticated Non-linear source activity and connectivity estimators has been overlooked (Patterson and Thompson, 1971, Harville, 1977; Friston et al., 2007; Wipf et al., 2009; Belardinelli et al., 2012; Wu et al., 2016; Valdes-Sosa, 1996, Bosch-Bayard, et al., 2001). These are precisely the points of this paper:

1- Presenting a family of both Source Activity and Connectivity Non-linear estimators, denominated here Brain Connectivity Variable Resolution Tomographic Analysis (BC-VARETA).
2- Introducing a highly optimized method for the connectivity estimation based on the local quadratic approximation of hermitic gaussian graphical models with penalization function of the LASSO family.

3- Proposing measures of the Type I and II Leakage distortion in the context of BC-VARETA, that will be generalizable to other Non-linear MEEG methods.

## 2 Methods

### 2.1 Bayesian Model of MEEG Sources Activity and Connectivity

For the MEEG recorded signals, in the Fourier space, the Forward Model at a single Frequency Component is expressed by the general equation below. Check Appendix for the mathematical notation and definition of variables across this manuscript.

$$\boldsymbol{v}_m = \mathbf{L}_{v\iota}\boldsymbol{\iota}_m + \boldsymbol{\xi}_m; m \in \mathbb{M} \qquad [2.1.1]$$

The vectors of MEEG measurements $\boldsymbol{v}_m$ and signal noise $\boldsymbol{\xi}_m$, are independent Random Samples $m \in \mathbb{M}$, defined on the p-size Scalp Sensors (Electrodes) Space $\mathbb{E}$, meanwhile the sources activity random vector $\boldsymbol{\iota}_m$ is defined on the q-size discretized Gray Matter Space $\mathbb{G}$. The p × q-size Source to Data Transfer Operator (SDTO) $\mathbf{L}$ (transformation of spaces $\mathbb{G} \to \mathbb{E}$) builds on a discretization of the Lead Field from a head conductivity model (Riera and Fuentes, 1998; Valdés-Hernández et al. 2009).

Construing a Model of MEEG source localization and connectivity, upon the equation [2.1.1], can be tackled in general by the Bayesian formalism (MacKay, 2003), which involves categorizing as random variables the MEEG Measurements (*Data*) $\boldsymbol{v}_m$ and Source Activity (*Parameters*) $\boldsymbol{\iota}_m$. The model builds on a Parametric representation of the signal noise $\boldsymbol{\xi}_m$ and sources activity $\boldsymbol{\iota}_m$ Probability Density Functions (*pdf*). It is commonly given by hierarchically conditioned Gaussian models, i.e. Multivariate Circularly Symmetric Complex Gaussians $N^{\mathbb{C}}$, of the *Data* Likelihood and *Parameters* Prior. The parametrization within these distributions introduces an additional category of random variables denominated (*Hyperparameters*) $\Xi$. See its schematic representation by More-Penrose diagrams in *Figure 1*.

***Likelihood***

$$\boldsymbol{v}_m|\boldsymbol{\iota}_m, \Xi \sim N_{\mathrm{p}}^{\mathbb{C}}(\boldsymbol{v}_m|\mathbf{L}_{v\iota}\boldsymbol{\iota}_m, \boldsymbol{\Sigma}_{\xi\xi}); m \in \mathbb{M} \qquad [2.1.2]$$

$$\boldsymbol{\Sigma}_{\xi\xi} = \sigma_\xi^2 \mathbf{R}_{\xi\xi}; \sigma_\xi^2 \in \Xi$$

***Parameters Prior***

$$\boldsymbol{\iota}_m|\Xi \sim N_{\mathrm{q}}^{\mathbb{C}}(\boldsymbol{\iota}_m|0, \boldsymbol{\Sigma}_{\iota}); m \in \mathbb{M} \qquad [2.1.3]$$

$$\boldsymbol{\Theta}_{\iota} = \boldsymbol{\Sigma}_{\iota}^{-1}; \boldsymbol{\Theta}_{\iota} \in \Xi$$

***Hyperparameters Priors***

$$\boldsymbol{\Theta}_{\iota} \sim exp(\Pi(\boldsymbol{\Theta}_{\iota}, \mathbf{A})|\alpha) \qquad [2.1.4]$$

$$\sigma_\xi^2 \sim exp(1/\sigma_\xi^2 \,|b) \qquad [2.1.5]$$

Above, $\Xi = \{\boldsymbol{\Theta}_{\iota}, \sigma_\xi^2\}$ represents the Model Parametrization (Hyperparameters). The Noise Covariance or *Data* GGM conditional Covariance matrix $\boldsymbol{\Sigma}_{\xi\xi}$ of formula [2.1.2], is assumed to be composed by a scalar random variable $\sigma_\xi^2$, representing the unknown nuisance Variance, and a known matrix $\mathbf{R}_{\xi\xi}$ (in the Cartesian space product $\mathbb{E} \times \mathbb{E}$) of the noise Covariance structure. The noise Covariance structure encodes information about the sensors correlated activity. These correlations are given either by shorting currents between adjacent electrodes' due to the scalp conductivity or common inputs from instrumentation/environmental noisy sources. The noise Precision (Variance) Exponential (Jeffry Improper) Gibbs *pdf* set up on, see formula [2.1.5], aims to bypass the nuisance level that could be assimilated into the **Parameters**. This is possible due to the monotonically increasing values of the noise Variance probability density assigned by the Jeffry Improper Prior, which allows for encoding the information about the noise inferior threshold into the parameter $b$.

The inverse of the Covariance matrix $\mathbf{\Sigma}_u$, Precision matrix $\mathbf{\Theta}_u$ (in the Cartesian space product $\mathbb{G} \times \mathbb{G}$), of the **Parameters** GGM, represents the source connectivity, see formula [2.1.3]. The general penalization function $\Pi$ at the argument of the exponential Prior in formula [2.1.4], imposes certain Structured Sparsity pattern on the connectivity. The Structured Sparsity can be encoded given information from the Gray Matter anatomical segmentation, by penalizing the groups of variables corresponding to the Gray Matter areas Intra/Inter-connections. The matrix $\mathbf{A}$ (in the Cartesian space product $\mathbb{G} \times \mathbb{G}$) within the General Penalization function, represents a probability mask of the anatomically plausible connections. The probability mask in case of the dense short-range connections, e.g. Intra-Cortical Connections, is defined as a deterministic spatially invariant empirical Kernel of the connections strength decay with distance. For the long-range connections, e.g. Inter-Cortical connection, it is given by probabilistic maps of the White Matter tracks connectivity strength from Diffusion Tensor Imaging (DTI). The global influence in the **Parameters** GGM of the connectivity Structured Sparsity penalization $\Pi(\mathbf{\Theta}_u, \mathbf{A})$ is controlled by the Scale Parameter (Regularization Parameter) $\alpha$, which can be fitted to the **Data** by means of some statistical criteria of goodness.

In general, the construction of the Model corresponds to the ubiquitous Bayesian representation of Linear State Space Models (LSSM), in both Time (Real) and Frequency (Complex) domain. The LSSM **Data** (Observation Equation) and **Parameters** (Autoregressive State Equation) are modeled by Multivariate Gaussian *pdf*s, whereas the Connectivity (Autoregression Coefficients Matrix) is represented by the Parameters' Precision Matrix, (Wills et al., 2009; Faes et al., 2012; Galka et al., 2004; Pascual-Marqui et al., 2014; Valdes-Sosa 2004; Lopes da Silva et al., 1980; Baccalá and Sameshima, 2001; Babiloni et al., 2005).

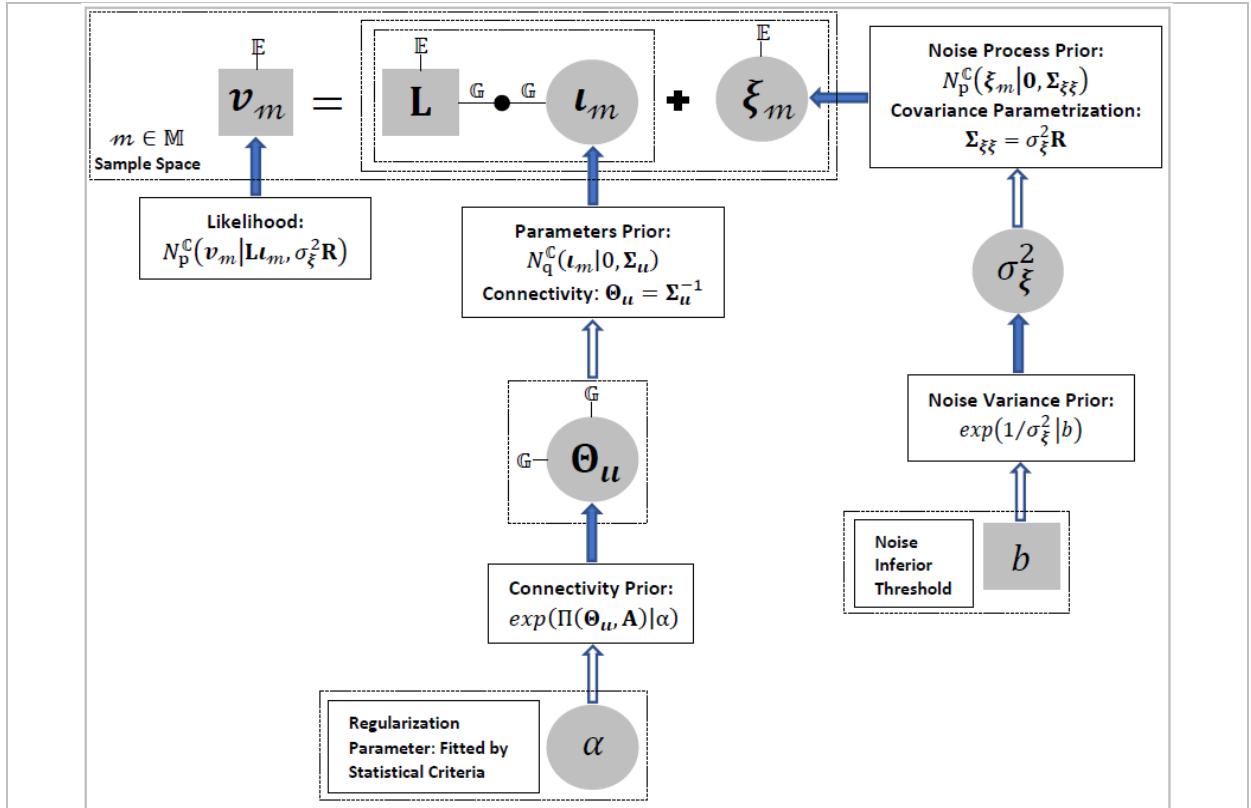

*Figure 1: More-Penrose diagram of the MEEG Source Activity and Connectivity Bayesian Model and its Priors. The model Variables and Prior knowledge are represented with gray circles and squares respectively. The filled arrows represent Random Variables generation by a specific pdf and the unfilled arrows the corresponding pdf parametrization.*

## 2.2 Type I Leakage and the Brain Connectivity Variable Resolution Tomographic Analysis

The Bayesian Model depicted in *Section 2.1* revendicates a large family of Linear, or Non-Linear/Hybrid iterated Sources Activity estimators, see *Table 1*. Along this family, the formulation of the source activity estimator, denominated First Level of Inference into the Bayesian Formalism, is also common to BC-VARETA. The estimators of this Model are given independently for each frequency component, since it does not consider Priors that could link the analysis along Frequency Domain. The First Level of Inference consists on maximizing the Multivariate Gaussian *pdf* derived from **Parameters**' Posterior Analysis. This is given upon fixed values of the **Hyperparameters** $\widehat{\Xi}^{(k)}$, within an outer cycle indexed $(k)$ of the **Parameters** and **Hyperparameters** iterated computation, see Section C of (Paz-Linares and Gonzalez-Moreira et al, 2018):

$$\iota_m | v_m, \widehat{\Xi}^{(k)} \sim N_q^{\mathbb{C}}\left(\iota_m \middle| \hat{\iota}_m^{(k)}, \breve{\Sigma}_\iota^{(k)}\right) \qquad [2.2.1]$$

The **Parameters**' Posterior Mean (Sources Activity iterated estimator) $\hat{\iota}_m^{(k)}$, given the Data $v_m$, is expressed through the iterated auxiliary quantities of the Data to Sources 'Transference Operator' $\breve{T}^{(k)}$ (transformation of spaces $\mathbb{E} \to \mathbb{G}$) and the **Parameters**' Posterior Covariance $\breve{\Sigma}_\iota^{(k)}$ (in the space $\mathbb{G} \times \mathbb{G}$). Both depending on the Hyperparameters iterated estimators $\widehat{\Xi}^{(k)}$.

$$\hat{\imath}_m^{(k)} \leftarrow \breve{\mathbf{T}}^{(k)} \boldsymbol{v}_m \qquad [2.2.2]$$

$$\breve{\mathbf{T}}^{(k)} = \breve{\boldsymbol{\Sigma}}_{\boldsymbol{u}}^{(k)} \mathbf{L}^{\mathcal{T}} \left( \hat{\sigma}_{\xi}^{2\,(k)} \mathbf{R} \right)^{-1} \qquad [2.2.3]$$

$$\breve{\boldsymbol{\Sigma}}_{\boldsymbol{u}}^{(k)} = \left( \mathbf{L}^{\mathcal{T}} \left( \hat{\sigma}_{\xi}^{2\,(k)} \mathbf{R} \right)^{-1} \mathbf{L} + \widehat{\boldsymbol{\Theta}}_{\boldsymbol{u}}^{(k)} \right)^{-1} \qquad [2.2.4]$$

*Table 1: Family of Linear/Non-Linear/Hybrid Source Activity estimators*

| | Parameters Covariance | Data Conditional Covariance | Algorithm | Linear Non-Linear Hybrid |
|---|---|---|---|---|
| **Minimum Norm Estimator (MME)** (Hämäläinen and Ilmoniemi, 1994) | Constrained to Scaled Identity Matrix. Fixed value. Connectivity-Postprocessing. | Constrained to Scaled Identity Matrix. Fixed value. | Explicit *Parameters* Posterior Mean estimator. | Linear |
| **Low Resolution Electromagnetic Tomography (LORETA)** (Pascual-Marqui et al., 1994). | Fixed Laplacian Operator. Connectivity-Postprocessing. | Full Matrix. Prior Free. | Iterated Explicit *Parameters* Posterior Mean estimator and empirical formulas of the Conditional *Data* Covariance. | Hybrid |
| **Exact Low Resolution Electromagnetic Tomography (eLORETA)** (Pascual-Marqui et al., 2006). | Constrained to a Diagonal Matrix. Prior Free. Connectivity-Postprocessing. | Full Matrix. Prior Free. | Iterated Explicit *Parameters* Posterior Mean estimator and empirical formulas of the Conditional *Data* Covariance and *Parameters* Variances. Provides Zero Localization Error in case of a single Source reconstruction. | Hybrid |
| **Standardized Low Resolution Electromagnetic Tomography (eLORETA)** (Pascual-Marqui, 2002). | Full Matrix (Connectivity). Prior Free. | Full Matrix. Prior Free. | Iterated Explicit *Parameters* Posterior Mean estimator and empirical formulas of the Conditional *Data* Covariance and *Parameters* Covariances. | Hybrid |
| **Variable Resolution Tomographic Analysis (VARETA)** (Valdes-Sosa, 1996, Bosch-Bayard, et al., 2001). | Full Matrix (Connectivity). Prior Free. | Constrained to Scaled Identity Matrix. Prior free. | Iterated Explicit *Parameters* Posterior Mean estimator and Expectation Maximization (EM) formulas of the Conditional *Data* Variance and *Parameters* Covariance. | Non-linear |
| **Automatic Relevance Determination (ARD)** (Neal, 1998; Tipping, 2001; Sato et al., 2004; Wipf et al., 2006; Wipf and Rao, 2007; Daunizeau and Friston, 2007). | Constrained to a Diagonal Matrix. Jeffrey Improper Priors. Connectivity-Postprocessing. | Constrained to Scaled Identity Matrix. Prior free. | Iterated Explicit *Parameters* Posterior Mean estimator and Empirical Bayes (EB) formulas of the Conditional *Data* Variance and *Parameters* Variances. Induces Sparsity by pruning the *Parameters* Variances estimates. | Non-linear |

| Restricted Likelihood Maximization (ReLM) (Patterson and Thompson, 1971, Harville, 1977; Friston et al., 2007; Wipf et al., 2009; Belardinelli et al., 2012; Wu et al., 2016) | Full Matrix (Connectivity) hyper-parametrized on Function Basis. Sparse Priors on the Function Basis *Hyperparameters*. | Constrained to Scaled Identity Matrix. Prior free. | Iterated Explicit *Parameters* Posterior Mean estimator and Maximum Likelihood plus Restrictions formulas of the Conditional *Data* Variance and Function Basis *Hyperparameters*. | Non-linear |
|---|---|---|---|---|
| **Structured Sparse Bayesian Learning (SSBL)** (Wipf et al., 2010; Zhang and Rao, 2011; Babacan et al., 2012; Wan et al., 2014; Balkan et al. 2014; Paz-Linares et al., 2017; Zhang et al., 2016; Zhang et al., 2017). | Constrained to a Diagonal Matrix (or Block Diagonal). Sparse Gamma Priors. Connectivity-Postprocessing. | Constrained to Scaled Identity Matrix. Jeffrey Improper Prior. | Iterated Explicit *Parameters* Posterior Mean estimator and Empirical Bayes (EB) formulas of the Conditional *Data* Variance and *Parameters* Variances. | Non-linear |

### *2.2.1 Measures of Type I (Activity) Leakage*

Evaluating the Type I Leakage given to Volume Conduction distortion itself should be done in an isolated point Source scenario. In the general context of Signal Processing this situation is represented by the concept of 'Point Spread Function' (**PSF**). In this case it is defined as the estimated Sources Activity from synthetic Data due to a single Unitary Source, at an arbitrary point $j_0$ of the Gray Matter space $\mathbb{G}$. The Activity of a single Unitary Active Source is mathematically represented by the Kronecker Delta, denoted by the vector $\delta(j_0)$:

$$\delta_j(j_0) = \begin{cases} 0, & j \neq j_0 \\ 1, & j = j_0 \end{cases} \qquad [2.2.5]$$

Computing the **PSF** involves the projection $\mathbb{E} \to \mathbb{G}$, by effectuating formulas [2.2.2], [2.2.3] and [2.2.4] till convergence, given inputted Data obtained from the projected Kronecker Delta $\mathbb{G} \to \mathbb{E}$ in [2.2.5], i.e. $lim_{k\to\infty}\left(\breve{\mathbf{T}}^{(k)}\mathbf{L}\delta(j_0)\right)$. Expressing the PSF by this formula alone constitutes an idealization of the real MEEG Source Localization scenario. A fair approach should consider avoiding the Inverse Crime (Kaipio & Somersalo 2004), which consists on modifying the Lead Field of Data generation from the one used within the Inference framework, and corrupting the Data with Noise, from all possible sources including biological, environmental and instrumentation, at acceptable levels.

In a situation as such, a suitable definition of **PSF**, will be given by the Variances of the iterated auxiliary quantity of Sources Activity Empirical Covariance (SEC) $\breve{\mathbf{S}}_u^{(k)}$ (in the Cartesian space product $\mathbb{G} \times \mathbb{G}$). An explicit and compact expression of the SEC can be attained by the projection $\mathbb{E} \times \mathbb{E} \to \mathbb{G} \times \mathbb{G}$ of the Data Empirical Covariance $\mathbf{S}_{vv}$ (in the Cartesian space product $\mathbb{E} \times \mathbb{E}$) through the Transference Operator $\breve{\mathbf{T}}^{(k)}$, see Section D of (Paz-Linares and Gonzalez-Moreira et al, 2018).

$$\breve{\mathbf{S}}_u^{(k)} \leftarrow \breve{\mathbf{T}}^{(k)}\mathbf{S}_{vv}\breve{\mathbf{T}}^{(k)\dagger} \qquad [2.2.6]$$

$$\mathbf{S}_{vv} = \frac{1}{m}\sum_{m=1}^{m} \boldsymbol{v}_m \boldsymbol{v}_m^\dagger \qquad [2.2.7]$$

The computation of **PSF** is effectuated on a synthetic $\mathbf{S}_{vv}^{sim}$, obtained from the projection $\mathbb{G} \times \mathbb{G} \to \mathbb{E} \times \mathbb{E}$, through an independent Lead Field $\mathbf{L}_{sim}$, of the Kronecker Delta $\delta(j_0)$ plus Noise samples, represented by the Real/Complex vectors $\xi_m$:

$$v_m^{sim} = \mathbf{L}_{sim}\delta(j_0) + \xi_m, m = 1 \ldots \mathrm{m} \quad [2.2.8]$$

The explicit formula of the PSF, denoted here as $\mathcal{P}$, is given by the diagonal values (Variances) of the SEC iterated estimator, in formula [2.2.6], after convergence of the outer cycle:

$$\mathcal{P}(j_0) \leftarrow \lim_{k \to \infty} \left( diag\left( \check{\mathbf{T}}^{(k)} \mathbf{S}_{vv}^{sim} \check{\mathbf{T}}^{(k)\dagger} \right) \right) \quad [2.2.9]$$

The **PSF** distortion can be evaluated in general by any Measure of its difference to Ground Truth, Kronecker delta in [2.2.5]. Particularly, for single point spreading like scenarios and when the Gray Matter space $\mathbb{G}$ is collapsed to a bidimensional Manifold, i.e. surfaces of different Brain structures, a measure universally adopted is the Spatial Dispersion (**SD**) according to the Geodesic Distance. In such scenario the Type I Leakage due to Volume Conduction distortion is expressed through the Spatial Dispersion of the Point Spread Function (**SD-PSF**). It is defined as the Standard Deviation of the Geodesic Distance $d_{jj_0}$ between pairs of points indexed $(j, j_0)$, for $j = 1 \ldots q$, in the Gray Matter space $\mathbb{G}$, with probability mass given by the absolute values of the PSF, denoted mathematically as $\mathbf{SD}_{\delta(j_0)}(\mathcal{P}(j_0))$, see formula below:

$$\mathbf{SD}_{\delta(j_0)}(\mathcal{P}(j_0)) = \sqrt{\frac{\sum_{j=1}^{q} d_{jj_0}^2 |\mathcal{P}_j(j_0)|}{\sum_{j=1}^{q} |\mathcal{P}_j(j_0)|}} \quad [2.2.10]$$

In a more general scenario, where the Data is given by a composition of multiple Unitary Active Sources $v_m^{sim} = \mathbf{L}_{sim}\delta(j_0, j_1, \cdots) + \xi_m$, $m = 1 \ldots \mathrm{m}$, the concept of **PSF** requires to be extended, i.e. 'Generalized Spread Function' **GSF**, denoted mathematically as $\mathcal{P}(j_0, j_1, \cdots)$. In this case the Type I Leakage is given by the composition of two distortive effects, i.e. the Volume Conduction and superposed Scalp projection of multiple Sources, which cannot be measured by simply using the **SD**. The Earth Movers' Distance (**EMD**) between the **GSF** and the Ground Truth (**EMD-GSF**) would suit as a measure representative of the distortion in this general scenario, denoted mathematically as $\mathbf{EMD}_{\delta(j_0, j_1, \cdots)}(\mathcal{P}(j_0, j_1, \cdots))$. In the State of the Art of Inverse Solution the **EMD** has been stablished as the most sensitive when compared to other quality measures, i.e. typical Dipole Localization Error or Binary Classification, i.e. Receiving Operating Characteristic, Precision, Recall and F1. See its definition in (Molins et al. 2008, Grova et al., 2006, Haufe et al., 2008). This concept is applicable in general to any definition of the Active sources, e.g. Vector made of patches with random extensions and random elements.

### 2.2.2 Second Level of Inference of the Brain Connectivity Variable Resolution Tomographic analysis and its influence on variable selection (Leakage)

Meanwhile, the First Level of Inference of the Methods described in *Table 1* constitutes an invariant, a distinct aspect was its Parametrization structure and Priors defined. This is definitory at the denominated Second Level of inference or estimation of Hyperparameters $\hat{\boldsymbol{\Xi}}^{(k)}$, which biases the variables selection into the iterated estimation scheme, and thus the amount of Leakage carried by the **Parameters** and **Hyperparameters**. This effect is determined, at the First Level of Inference, by the Resolution (sparsity) in Variable Selection of the Transference Operator $\check{\mathbf{T}}^{(k)}$, which is influenced by the scale and/or degree of sparsity of the Precision matrix $\hat{\boldsymbol{\Theta}}_u^{(k)}$ and Data Nuisance $\hat{\sigma}_\xi^{2(k)}$ estimators. Those transitively influence the

Resolution through its balance into the Parameters' Posterior Covariance Matrix $\breve{\Sigma}_u^{(k)}$, see formulas [2.2.3], [2.2.4], [2.2.6] and [2.2.7]. The Resolution in the estimation will be globally determined by two interacting elements. First: The choices of Hyperparameters Posterior analysis strategies, i.e. EB, EM, ARD, ReLM, etc. Second: The bias introduced by the **Hyperparameters** Priors, i.e. ad hoc structure of the Data Conditional Covariance **R**, Prior *pdf*s on the Sources' Activity Precisions matrix $\Theta_u$ and Data Nuisance Variance $\sigma_\xi^2$. Thus, the essential constituent of the BC-VARETA methodology is the definition of the Priors and Inference strategy at the Second Level along with adequate statistical guarantees.

The EM algorithm (Dempster et al., 1977; Liu and Rubin, 1994; McLachlan and Krishnan, 2007) constitutes an explicit way to tackle the **Hyperparameters** estimation. This is done by iteratively maximizing its approximated representation of the intractable Data Type II likelihood $p(\{v_m\}_{m=1}^m|\Xi)$, by the so-called **Data** Expected Log-Likelihood $Q(\Xi, \widehat{\Xi}^{(k)})$. The **Data** Expected Log-Likelihood is construed by the marginalization (Expectation) of the **Data** and **Parameters** Joint *pdf* $log(p(\{v_m\}_{m=1}^m, \{\iota_m\}_{m=1}^m|\Xi))$, given through formulas [2.1.2] and [2.1.3], by the parameters Posterior *pdf* $p(\iota_m|v_m, \widehat{\Xi}^{(k)})$, see formula [2.2.1]. The BC-VARETA methodology, meshed to the EM scheme at the Second Level of Inference, constitutes a special case of **Hyperparameters** Penalized Posterior Analysis. It is defined as the maximization of the approximated **Hyperparameter**'s Posterior *pdf*, given by the combination the **Data** Expected Likelihood and **Hyperparameter**'s Priors, see Section D of (Paz-Linares and Gonzalez-Moreira et al, 2018):

$$\Xi|\{v_m\}_{m=1}^m, \widehat{\Xi}^{(k)} \sim e^{Q(\Xi, \widehat{\Xi}^{(k)})} p(\Xi) \qquad [2.2.11]$$

The **Data** Expected Log-Likelihood has a close form expression on the **Hyperparameters**, given the Data Empirical Covariance and the iterated estimators of the Data to Sources Transference Operator, Sources Posterior Covariance, and an iterative auxiliary quantity denominated Effective Sources Empirical Covariance (ESEC) $\breve{\Psi}_u^{(k)}$:

$$Q(\Xi, \widehat{\Xi}^{(k)}) = -mp\, log(\sigma_e^2) - (m/\sigma_e^2)\, tr\left(\left(I_p - L\breve{T}^{(k)}\right)^\dagger R^{-1}\left(I_p - L\breve{T}^{(k)}\right) S_{vv}\right) \cdots$$

$$-(m/\sigma_e^2) tr\left(L^T R^{-1} L \breve{\Sigma}_u^{(k)}\right) + m\, log|\Theta_u| - m\, tr\left(\Theta_u \breve{\Psi}_u^{(k)}\right) \qquad [2.2.12]$$

$$\breve{\Psi}_u^{(k)} = \breve{\Sigma}_u^{(k)} + \breve{S}_u^{(k)} \qquad [2.2.13]$$

The Data Expected Log-Likelihood in formula [2.2.12] is a Concave function on $\Xi$, but the Concavity of its associated Posterior *pdf* in formula [2.2.11] can be only ensured with a pertinent definition of the Priors. In addition, the approximated **Hyperparameter**'s Posterior analysis of EM algorithm only guarantees reaching a local maximum of the actual **Hyperparameters**' Posterior, i.e. $p(\Xi|\{v_m\}_{m=1}^m) \propto p(\{v_m\}_{m=1}^m|\Xi)p(\Xi)$. The proximity of this local maximum to the global maximum is determined also by the selection of the Priors. The Gibbs Priors of formulas [2.1.5] and [2.1.6] guarantee the Concavity whenever the exponent arguments redress the mathematical definition of norm, i.e. a non-negative scalar function that satisfies 1) triangle inequality, 2) absolutely-scalable and 3) positive-definite.

The Precision Matrix estimation, given by Posterior Analysis of equation [2.2.11], is expressed through the minimization of a Target Function that resembles the structure of an equivalent Sources GGM with Effective Sources Empirical Covariance (ESEC) Matrix $\breve{\Psi}_u^{(k)}$. Under the convention $\alpha = \lambda m$, where the $\lambda$ can be interpreted as the GGM effective Regularization Parameter, the Precision matrix estimator is given by the following formula, see Section E of (Paz-Linares and Gonzalez-Moreira et al, 2018):

$$\widehat{\Theta}_u^{(k+1)} \leftarrow argmin_{\Theta_u} \left\{ -\log|\Theta_u| + tr\left(\Theta_u \breve{\Psi}_u^{(k)}\right) + \lambda \Pi(\Theta_u, \mathbf{A}) \right\} \quad [2.2.14]$$

Setting up Sparse Models as General Penalty Function has been stablished in similar scenarios of Variable Selection, i.e. Graphical Models estimation (Jordan, 1998; Attias, 2000; Friedman et al, 2008; Mazumder et al. 2012; Wang, 2012; Wang, 2014; Schmidt, 2010; Hsieh, 2014; Danaher et al., 2014; Zhang and Zou, 2014; Yuan and Zheng, 2017; Drton and Maathuis, 2017). Some of the most common Penalty Functions are referred into the family of Graphical LASSO Models, see *Table 2* below.

*Table 2: Graphical LASSO family Penalty Functions Models*

|  | Penalty function $\Pi(\Theta_u, \mathbf{A})$ |  |
|---|---|---|
| **Graphical LASSO (GLASSO)** | $\|\Theta_u\|_{1,\mathbf{A}}$ |  |
| **Graphical Elastic Net (GENET)** | $\|\Theta_u\|_{1,\mathbf{A}_1} + \|\Theta_u\|_{2,\mathbf{A}_2}^2$ |  |
| **Graphical Group Lasso (GGLASSO)** | $\sum_{j=1}^{\mathbb{q}} \left\|\Theta_{u_{(\mathbb{K}_j)}}\right\|_{2,\mathbf{A}_{(\mathbb{K}_j)}}$ | ; $\mathbb{K}_j \subset \mathbb{G} \times \mathbb{G}; j = 1 \cdots n$ |

Nevertheless, as for choosing the Penalty Function and Regularization Parameter there is not ubiquitous rule. It is usually assumed that, for a given Penalty Function, fitting the Regularization Parameter by some Statistical Criteria by would suffice to rule out the ambiguity on the Variable Selection sparsity level (Resolution). This approach does not provide a Statistical guarantee, as discussed in (Jankova and Van De Geer, 2015, 2017), due the biasing introduced in the estimation by the Sparse Penalty in any case. For the typical Graphical LASSO, a solution was recently presented in (Jankova and Van De Geer, 2018) through an unbiased Precision Matrix estimator $(\widehat{\Theta}_u)_{unbiased}^{(k+1)}$.

$$\left(\widehat{\Theta}_u\right)_{unbiased}^{(k+1)} = 2\widehat{\Theta}_u^{(k+1)} - \widehat{\Theta}_u^{(k+1)} \breve{\Psi}_u^{(k)} \widehat{\Theta}_u^{(k+1)} \quad [2.2.15]$$

For the conditions $\Pi(\Theta_u, \mathbf{A}) = \|\Theta_u\|_1$ and $\lambda = \sqrt{\log(q)/m}$, it is demonstrated, for the elements into the unbiased estimator of formula [2.2.5] $\left(\left(\widehat{\Theta}_u\right)_{unbiased}^{(k+1)}\right)_{ij}$, a tendency to the Model Precision Matrix elements $(\Theta_u)_{ij}$ with Complex Normal *pdf* of consistent variances $\sigma_{ij}\left(\left(\widehat{\Theta}_u\right)_{unbiased}^{(k+1)}\right) = \left(\widehat{\Theta}_u^{(k+1)}\right)_{ii} \left(\widehat{\Theta}_u^{(k+1)}\right)_{jj} + \left(\widehat{\Theta}_u^{(k+1)}\right)_{ij}$ rated by $\sqrt{m}$:

$$\left(\left(\widehat{\Theta}_u\right)_{unbiased}^{(k+1)}\right)_{ij} \sim N_1^{\mathbb{C}} \left(\left(\left(\widehat{\Theta}_u\right)_{unbiased}^{(k+1)}\right)_{ij} \middle| (\Theta_u)_{ij}, \frac{\sigma_{ij}\left(\left(\widehat{\Theta}_u\right)_{unbiased}^{(k+1)}\right)}{\sqrt{m}}\right) \quad [2.2.16]$$

For the Nuisance *Hyperparameter* a close form estimator can be obtained, by the unique zero of the resulting equation given the derivative $\frac{\partial}{\partial \sigma_e^2}$ of formula [2.2.11]. It is expressed in below, by reformulating the Jeffry Improper Prior *pdf* Rate Parameter as $b = mp\epsilon$, where $\epsilon$ is the Data Nuisance effective inferior threshold, see Section E of (Paz-Linares and Gonzalez-Moreira et al, 2018):

$$\hat{\sigma}_\xi^{2(k+1)} \leftarrow \frac{tr\left(\left(\mathbf{I}_p - \mathbf{L}\breve{\mathbf{T}}^{(k)}\right)^\dagger \mathbf{R}^{-1}\left(\mathbf{I}_p - \mathbf{L}\breve{\mathbf{T}}^{(k)}\right)\mathbf{S}_{vv}\right)}{p} + \frac{tr\left(\mathbf{L}^T\mathbf{R}^{-1}\mathbf{L}\breve{\mathbf{\Sigma}}_u^{(k)}\right)}{p} + \epsilon \quad [2.2.17]$$

With equation [2.2.17] we attain a formulation in which the biasing given the Nuisance *Hyperparameter* Prior relies on the Data Nuisance effective inferior limit $\epsilon$ and the Noise Covariance Structure $\mathbf{R}$, which are quantities that can be experimentally informed. The interpretability of this Prior structure would thus rule out any ambiguity on its choice.

## 2.3 Estimation of the MEEG Sources Gaussian Graphical Model

Despite the growing interest on the GGM's given its applicability in several fields, drawbacks of the State of Art methodologies prevent of utilizing them in the scenario of Electrophysiological Sources Localization and Connectivity, we mention some of them in *Table 3* below.

| *Table 3: Drawbacks of the Gaussian Graphical Models methodologies* | |
|---|---|
| **Complex Variable** | The stablished algorithms do not deal with Complex Variables, limiting their implementation on the Frequency Domain Connectivity Analysis. |
| **Stability** | They are based on very unstable strategies, such as Coordinate Updates of the Target Function Descend Direction or Alternating Direction Methods of Multipliers. |
| **Dimensions** | The High Dimensionality (common scenario in Brain Connectivity) along with the combined effect of instability and Inverses computation, at every iteration of the Coordinate Updates cycle, constitutes an important reason for these algorithms frequent crashing. |
| **Bayesian analysis** | Addressing this problem from the perspective of Machine Learning or Optimization Theory is the common trend to most of Data Analysis groups, while a complete Bayesian insight to the structure and properties of the GGM and its Precision matrix Priors is still missing in State of the Art. |

Here we propose a revindication of the MEEG SGGM from the Bayesian perspective, that allows for obtaining a more general class of explicit Precision matrix (Connectivity) estimators. This is done by considering invariance properties of the GGM Wishart Likelihood and the hierarchical representation of the GGM Gibbs Priors'. The analogous **SGGM** representation of the Precision matrix Expected Posterior **pdf** can be expressed by a Wishart Likelihood $W^{\mathbb{C}}$ (Real/Complex) on the ESEC matrix $\breve{\mathbf{\Psi}}_u^{(k)}$ of m degrees of freedom and scale matrix $(m\mathbf{\Theta}_u)^{-1}$, combined with the Gibbs Prior **pdf** of formula [2.1.5].

$$\breve{\mathbf{\Psi}}_u^{(k)}\big|\mathbf{\Theta}_u \sim W_{\mathrm{q}}^{\mathbb{C}}\left(\breve{\mathbf{\Psi}}_u^{(k)}\big|(m\mathbf{\Theta}_u)^{-1}, m\right) \qquad [2.3.1]$$

$$\mathbf{\Theta}_u \sim e^{-\alpha \Pi(\mathbf{\Theta}_u)} \qquad [2.3.2]$$

$$W_{\mathrm{q}}^{\mathbb{C}}\left(\breve{\mathbf{\Psi}}_u^{(k)}\big|(m\mathbf{\Theta}_u)^{-1}, m\right) = \left|\breve{\mathbf{\Psi}}_u^{(k)}\right|^{(m-q)} |\mathbf{\Theta}_u|^m e^{-m\,tr\left(\mathbf{\Theta}_u \breve{\mathbf{\Psi}}_u^{(k)}\right)} \qquad [2.3.3]$$

The rigorous theoretical derivation of the strategy for the minimization of the Sources GGM Target Function is presented in Section G of (Paz-Linares and Gonzalez-Moreira et al, 2018). It is done by a modified Model through the hierarchical representation (in Complex Variable) of the exponential Priors by mixtures of Gaussian and Gamma **pdf** 's (Andrews and Mallows, 1974; Tipping, 2001; Schmolck and Everson, 2007; Faul and Tipping, 2002; Li and Lin, 2010; Kyung et al., 2010). With this hierarchical model it is derived a concave Local Quadratic Approximation (LQA) estimation strategy of the SGGM (Fan and Li, 2001; Valdés-Sosa et al., 2006; Sánchez-Bornot et al., 2008). We reformulate the LQA Target Function of the Precision Matrix into simple Quadratic Model, due to the Standardization GGM's Wishart Likelihood (Srivastava, 1965; Drton et al., 2008). The explicit Connectivity estimator of the Standard Quadratic Model is expressed as the unique solution of a Matrix Riccati equation (Lim, 2006; Honorio and Jaakkola, 2013).

## 2.4 Type II Leakage and the Brain Connectivity Variable Resolution Tomographic Analysis

The whole estimation strategy consists on the computation ESEC $\breve{\mathbf{\Psi}}_u^{(k)}$ with unbiased Precision Matrix $\left(\widehat{\mathbf{\Theta}}_u\right)_{unbiased}^{(k+1)}$, at the outer cycle indexed $k$-th. The unbiased Precision Matrix, given in formula [2.2.15], is computed from its SGGM LQA estimator, after the convergence of the $\widehat{\mathbf{\Theta}}_u^{(k,l)}$ and $\widehat{\mathbf{\Gamma}}^{(k,l)}$, given within an inner cycle indexed $l$-th, see Section H of (Paz-Linares and Gonzalez-Moreira et al, 2018):

$$\widehat{\mathbf{\Theta}}_u^{(k,l+1)} \leftarrow \frac{1}{2\lambda} \widehat{\mathbf{\Gamma}}^{(k,l)} \odot \left( \sqrt{\left(\left(\breve{\mathbf{\Psi}}_u^{(k)}\right)^{-1} \oslash \widehat{\mathbf{\Gamma}}^{(k,l)}\right)^{-2} + 4\lambda \mathbf{I}_q} - \left(\breve{\mathbf{\Psi}}_u^{(k)^{-1}} \oslash \widehat{\mathbf{\Gamma}}^{(k,l)}\right)^{-1} \right) \quad [2.4.1]$$

$$\widehat{\mathbf{\Gamma}}^{(k,l)} \leftarrow \left(-\mathbf{1}_q + \left(\mathbf{1}_q + 4(\lambda m)^2 \mathbf{A}^{\cdot 2} \odot abs\left(\widehat{\mathbf{\Theta}}_u^{(k,l)}\right)^{\cdot 2}\right)^{\frac{1}{2}}\right)^{\cdot \frac{1}{2}} \oslash \left(2^{\frac{1}{2}}(\lambda m)^{\frac{1}{2}} \mathbf{A}\right) \quad [2.4.2]$$

As in Section 2.3, the elementwise matrix operations in the notation introduce: $\odot$ as Hadamard product, $\oslash$ as Hadamard division, $abs(\ )$ as elementwise matrix absolute value, $(\ )^{\cdot 2}$ (elementwise matrix Square exponentiation), $(\ )^{\frac{1}{2}}$ (elementwise matrix Square Root), $\mathbf{I}_q$ (q × q identity matrix), $\mathbf{1}_q$ (q × q matrix of ones).

### 2.4.1 Measures of Type II (Connectivity) Leakage

Evaluating the Type II Leakage given to Volume Conduction distortion itself should be done in a scenario where a single connection (pair connected sources) is present. For the Connectivity the **PSF** definition admits an extension to the Cartesian spaces product of Cortical Manifolds $\mathbb{G} \times \mathbb{G}$, i.e. 'Cartesian Point Spread Function' (**CPSF**). It is defined as the Precision matrix estimator from synthetic Data due to a pair of cortical sources at arbitrary points $j_0$ and $j_1$ with Unitary Connectivity (Precision matrix). The Unitary Precision matrix of the pair of connected Sources is denoted here with the matrix $\mathbf{\Delta}(j_0, j_1)$ on the space $\mathbb{G} \times \mathbb{G}$, given by an outer product of Kronecker Delta functions:

$$\mathbf{\Delta}(j_0, j_1) = \left(\delta(j_0) + \delta(j_1)\right)\left(\delta(j_0) + \delta(j_1)\right)^{\mathcal{T}} \quad [2.4.3]$$

In analogy to the construction of the **PSF** in Section 2.2.1, the computation of the **CPSF** is effectuated on a synthetic $\mathbf{S}_{vv}^{sim}$, obtained from the projection $\mathbb{G} \times \mathbb{G} \to \mathbb{E} \times \mathbb{E}$, through an independent Lead Field $\mathbf{L}_{sim}$ (avoiding inverse Crime) of the simulated Cortical Sources Activity plus Noise samples (corruption of Data), represented by the Real/Complex vectors $\xi_m$. The simulated Sources Activity, represented by the Real/Complex vectors $\iota_m$, is taken from random samples of Gaussian Random Generator with Covariance matrix $(\mathbf{\Delta}(j_0, j_1))^+$, given by the Pseudoinverse operation over the nonzero block of $\mathbf{\Delta}(j_0, j_1)$:

$$\mathbf{S}_{vv}^{sim} = \frac{1}{m} \sum_{m=1}^{m} \mathbf{v}_m^{sim} \mathbf{v}_m^{sim\dagger} \quad [2.4.4]$$

$$\mathbf{v}_m^{sim} = \mathbf{L}_{sim} \iota_m + \xi_m, m = 1 \dots m \quad [2.4.5]$$

$$\iota_m \sim N_q\left(\iota_m \middle| \mathbf{0}, (\mathbf{\Delta}(j_0, j_1))^+\right), m = 1 \dots m \quad [2.4.6]$$

The Precision matrix iterated estimator of formula [2.4.1] comprises the ESEC $\breve{\mathbf{\Psi}}_u^{(k)}$, expressed by substituting into formula [2.2.13] the SEC formula [2.2.6] computed for synthetic Data Empirical

Covariance from [2.4.5]. Expressing $\breve{\Sigma}_u^{(k)}$ and $\breve{T}^{(k)}$ into [2.2.13] through the unbiased Precision Matrix $(\widehat{\Theta}_u)_{unbiased}^{(k+1)}$ we obtain:

$$\breve{\Psi}_u^{(k)} = \left(L^{\mathcal{T}}\left(\hat{\sigma}_\xi^{2(k)}R\right)^{-1}L + (\widehat{\Theta}_u)_{unbiased}^{(k+1)}\right)^{-1} \dots$$

$$\times \left(I_q + \frac{1}{\left(\hat{\sigma}_\xi^{2(k)}\right)^2} L^{\mathcal{T}} R^{-1} S_{vv}^{sim} R^{-1\dagger} L \left(\left(L^{\mathcal{T}}\left(\hat{\sigma}_\xi^{2(k)}R\right)^{-1}L + (\widehat{\Theta}_u)_{unbiased}^{(k+1)}\right)^{-1}\right)^{\dagger}\right) \quad [2.4.7]$$

The explicit formula of the **CPSF** is given by the Precision matrix iterated estimator of formula [2.4.1] with the ESEC defined in formula [2.4.7] after convergence of the inner cycle (indexed $l$) and the outer cycle (indexed $k$):

$$\mathcal{C}(j_0, j_1) \leftarrow$$

$$\frac{1}{\lambda} \lim_{\substack{l \to \infty \\ k \to \infty}} \left( \widehat{\Gamma}^{(k,l)} \odot \left( \sqrt{\left(\breve{\Psi}_u^{(k)^{-1}} \oslash \widehat{\Gamma}^{(k,l)}\right)^{-2} + 4\lambda I_q} - \left(\breve{\Psi}_u^{(k)^{-1}} \oslash \widehat{\Gamma}^{(k,l)}\right)^{-1}\right)\right) \dots$$

$$-\frac{1}{4\lambda^2} \lim_{\substack{l \to \infty \\ k \to \infty}} \left(\left(\widehat{\Gamma}^{(k,l)} \odot \left(\sqrt{\left(\breve{\Psi}_u^{(k)^{-1}} \oslash \widehat{\Gamma}^{(k,l)}\right)^{-2} + 4\lambda I_q} - \left(\breve{\Psi}_u^{(k)^{-1}} \oslash \widehat{\Gamma}^{(k,l)}\right)^{-1}\right)\right) \dots \right.$$

$$\left. \times \breve{\Psi}_u^{(k)} \left(\widehat{\Gamma}^{(k,l)} \odot \left(\sqrt{\left(\breve{\Psi}_u^{(k)^{-1}} \oslash \widehat{\Gamma}^{(k,l)}\right)^{-2} + 4\lambda I_q} - \left(\breve{\Psi}_u^{(k)^{-1}} \oslash \widehat{\Gamma}^{(k,l)}\right)^{-1}\right)\right)\right) \quad [2.4.8]$$

The **CPSF** distortion (difference to the Ground Truth of the Precision matrix in [2.4.3]) can be evaluated by considering a generalization of the **SD** measure to Connectivity. Its expression can be directly deduced from the Natural extension of Geodesic Distance Natural to pairs of points belonging to the Cartesian product of Cortical Manifolds $\mathbb{G} \times \mathbb{G}$. Analogously to the **SD-PSF** formulation of Section 2.2.1, the Spatial Dispersion of the Cartesian Point Spread Function (**SD-CPSF**), denoted mathematically as $SD_{\mathcal{C}(j_0,j_1)}$, is defined as follows:

"The Standard Deviation of the Cartesian Geodesic Distance $D_{(j,j')(j_0,j_1)}$ between pairs of points indexed $\{(j,j'), (j_0, j_1)\}$, for $j, j' = 1 \cdots q$ in the Cartesian product $\mathbb{G} \times \mathbb{G}$ of Cortical Manifolds, with probability mass built on the superior triangle of the **CSF**" (see its schematic representation in *Figure 2*).

$$SD_{\Delta(j_0,j_1)}(\mathcal{C}(j_0,j_1)) = \sqrt{\frac{\sum_{j=1}^{q}\sum_{j'=j+1}^{q} D_{(j,j')(j_0,j_1)}^2 |c_{jj'}(j_0,j_1)|}{\sum_{j=1}^{q}\sum_{j'=j+1}^{q} |c_{jj'}(j_0,j_1)|}} \quad [2.4.9]$$

$$D_{(j,j')(j_0,j_1)} = \sqrt{d_{jj_0}^2 + d_{j'j_1}^2} \quad [2.4.10]$$

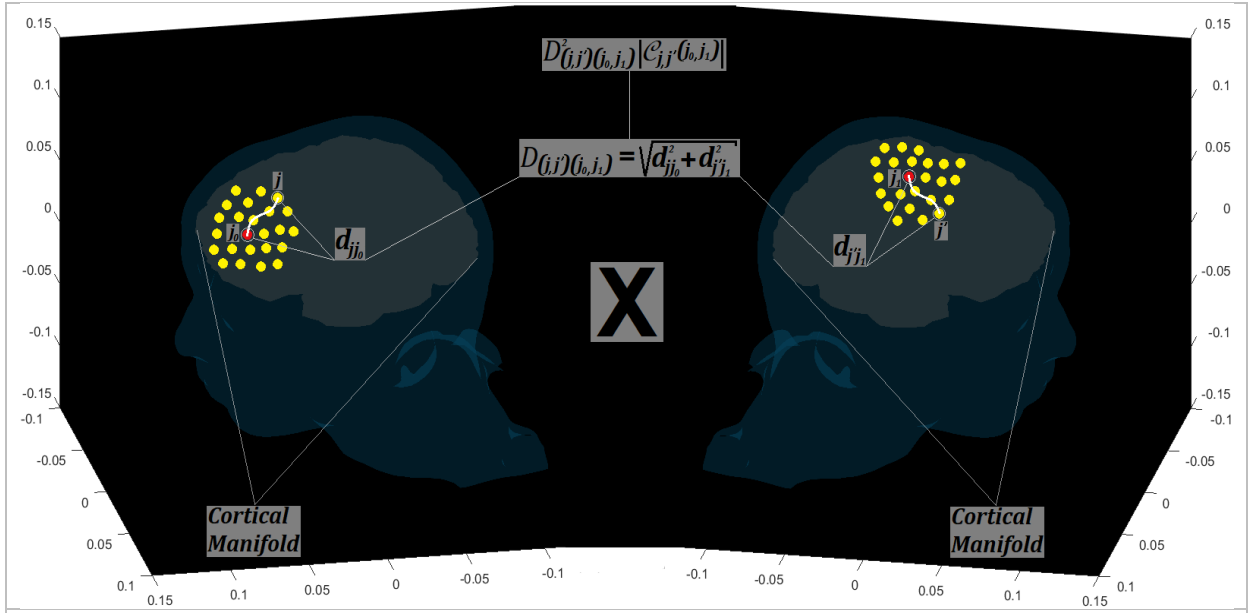

*Figure 2: Schematic representation of the **SD-CPSF** at the product of Cortical Manifold spaces given two elements. 1. The Cartesian Geodesic Distance $D_{(j,j')(j_0,j_1)}$ between the pair of Connected generators $(j,j')$, among the generators in the Non-zero elements of the **CPSF** superior triangle (Yellow Circles), and the pair of generators $(j_0,j_1)$ (Red Circles), at the centers of the of the Kronecker Deltas $\delta(j_0)$ and $\delta(j_1)$. 2. The corresponding contribution to the **SD** of the Cartesian Geodesic Distance weighted by the **CPSF** $D^2_{(j,j')(j_0,j_1)}|\mathcal{C}_{jj'}(j_0,j_1)|$.*

In similitude to what was discussed on the **PSF** and **GSF**, also the **CPSF** requires an extended representation for a general scenario in which the Data is given by a Unitary Precision Matrix given the composition of multiple Active Sources, i.e. $\Delta(j_0, j_1, \cdots) = (\delta(j_0) + \delta(j_1) + \cdots)(\delta(j_0) + \delta(j_1) + \cdots)^{\mathcal{T}}$. We denominate this representation "Cartesian Generalized Spread Function" **CGSF**, denoted mathematically as $\mathcal{C}(j_0, j_1, \cdots)$. Consequently, the **CGSP** distortion cannot be measured by using the **SD**. For this we consider the generalization to the Cartesian spaces product of Cortical Manifolds of the **EMD** measure (**EMD-CGSF**), denoted mathematically $EMD_{\Delta(j_0,j_1,\cdots)}(\mathcal{C}(j_0, j_1, \cdots))$, defined as follows:

"The sum of the typical **EMD** between all rows (columns) of the **CGSF** and Gold Standard, i.e. the projections in the cortical Manifold space $\mathbb{G}$ $\mathcal{C}_{j,:}(j_0, j_1, \cdots)$ ($\mathcal{C}_{:,j'}(j_0, j_1, \cdots)$) and $\Delta_{j,:}(j_0, j_1, \cdots)$ ($\Delta_{:,j'}(j_0, j_1, \cdots)$) for all $j = 1 \cdots q$ ($j' = 1 \cdots q$) in the Cortical Manifold space $\mathbb{G}$. The Complex rows/columns **EMD** is given by the sum of the EMD between its corresponding Real and Imaginary part."

$$EMD_{\Delta(j_0,j_1,\cdots)}(\mathcal{C}(j_0, j_1, \cdots)) = \sum_{j=1}^{q} EMD_{\Delta_{j,:}(j_0,j_1,\cdots)}(\mathcal{C}_{j,:}(j_0, j_1, \cdots)) \qquad [2.4.11]$$

$$EMD_{\Delta(j_0,j_1,\cdots)}(\mathcal{C}(j_0, j_1, \cdots)) = \sum_{j'=1}^{q} EMD_{\Delta_{:,j'}(j_0,j_1,\cdots)}(\mathcal{C}_{:,j'}(j_0, j_1, \cdots)) \qquad [2.4.12]$$

This concept is applicable in general to any definition of the sources Precision Matrix, e.g. Hermitian Matrix made of blocks with random extensions and random Complex elements.

# 3 Results

## 3.1 Simulation substrate

We evaluate the proposed estimators of Sources Localization and Connectivity on simulated EEG data. The simulation substrate was set up on a Cortical Manifold space $\mathbb{G}$, defined as 15K points Surface of the Gray Matter, with coordinates on the MNI Brain template (http://www.bic.mni.mcgill.ca). The Scalp Sensors space $\mathbb{E}$ was built on 343 electrodes, within 10-5 EEG Sensors system (Oostenveld and Praamstra, 2001). The Lead Fields, for both Simulations $\mathbf{L}_{sim}$ and for Reconstruction $\mathbf{L}$, were computed by BEM integration method accounting for a model of 5 head compartments (gray matter, cerebrospinal fluid, inner skull, outer skull, scalp) (Fuchs et al., 2002; Valdés-Hernández et al., 2009).

To avoid the Inverse Crime two individual subject Head Models were extracted from the corresponding T1 MRI images. The Electrophysiological Noise was defined by a composition of Sensors Noise and 500 Noisy Cortical Sources (approximating a 3% of the 15K points of the Cortical Manifold space $\mathbb{G}$), projected to the Scalp Sensors space $\mathbb{E}$. For practical computational limitations the Reconstruction was compute on a 6K points of Cortical Manifold space $\mathbb{G}$, obtained as a reduction from the original 15K points Surface of the Gray Matter. We design three kinds of characteristic simulations, to evaluate under different conditions the Leakage Effect in Source Localization (Type I) and Connectivity (Type II), see *Table 4*.

*Table 4:* Description of Simulations

| | |
|---|---|
| **Simulation 1** | |
| **Configuration** | A Unitary Source was placed at 500 random locations (following the procedure described in Section 2.2.1), with amplitude defined by the Kronecker Delta of formula [2.2.5]. 400 EEG Data trials were created at the Scalp Sensors space, by adding Noise Samples at 5dB Level to each projected Unitary Source. |
| **Aims** | Evaluate the Type I Leakage given to Volume Conduction distortion in an isolated point Source scenario, by the ***SD-PSF*** and ***EMD-PSF***. |
| **Simulation 2** | |
| **Configuration** | Two Sources were placed in 500 random configurations (following the procedure described in Section 2.4.1). 400 EEG Data trials were created, at the Scalp Sensors space, by projecting Samples from a Gaussian Random Generator given for each configuration. The Gaussian Random Generator Covariance structure was defined as the Inverse of the Unitary Precision matrix (Connectivity), given in formula [2.4.3]. Also, 5dB Level Noise were added to the Data of each projected Sample. |
| **Aims** | Evaluate the Type I Leakage given the composition the distortive effects of Volume Conduction and two Sources superposed at the Scalp projection, by the ***EMD-GSF***. <br> Evaluate the Type II Leakage given to Volume Conduction distortion in a scenario where a single connection is present, by the ***SD-CPSF*** and ***EMD-CPSF***. |
| **Simulation 3** | |
| **Configuration** | Four Sources were placed in 500 random configurations where only thwo of them were connected. 400 EEG Data trials were created, at the Scalp Sensors space following the same procedure as in Simulation 2. |
| **Aims** | Evaluate the Type I Leakage given the composition the distortive effects of Volume Conduction and multiple Sources superposed at the Scalp projection, by the ***EMD-GSF***. <br> Evaluate the Type II Leakage given to Volume Conduction distortion in a scenario where multiple connections are present, by the ***EMD-CGSF***. |

## 3.3 Validation Methods and software platforms

For reproducibility of the entire methodology proposed in this work results we provide a complementary routines package, i.e. Sources Activity and Connectivity estimators along with the *Type I* and *II* Leakage measures **SD-PSF**, **EMD-GSF**, **SD-CPSF** and **EMD-CGSP**, publicly available in MATLAB format at the GitHub link: https://github.com/dpazlinares/BC-VARETA. The evaluation of **SD-PSF**, **EMD-GSF**, **SD-CPSF** and **EMD-CGSP** measures was also extended to different Methods within the State-of-the-Art Electrophysiological Source analysis, implemented into the FIELDTRIP software package publicly available at: https://github.com/fieldtrip/fieldtrip/blob/master/ft_sourceanalysis.m.

For comparison purpose we selected the Exact version of LORETA (eLORETA) (Pascual, 2002), which constitutes the most stablished and robust (under a wide range of conditions), among the of family Source Activity estimators described in **Table 1**. The eLORETA **SD-PSF** and **EMD-GSF** were computed from the SEC at the Fist Level of Inference after convergence, according to the theory in *Subsection 2.2.1*. The **SD-CPSF** and **EMD-CGSF** was taken from the Inverse of the Source Covariance ma Enpirical Formula matrix at the Second Level of Inference after convergence.

We also consider the Linearly Constrained Minimum Variance (LCMV) (Van Veen et al. 1997), well stablished among the family of Beam Former methods. The LCMV constitutes a qualitative different approach in comparison with the family of iterated Source Activity estimators described in **Section 2.2**, that enriches our validation with a higher contrast of the results. Roughly, it consists on the Spatial Filtering of the Forward Equation [2.1.1], under similar assumptions of the Noise and Sources Activity *pdf*'s and in **Subsection 2.1**. The LCMV focuses only in the Sources' Variances, ruling out the Covariance structure from the First Level of Inference, but it also provides a Connectivity analysis through empirical formulas of the Sources' Covariance. For the computation of the **SD-PSF**, **EMD-GSF**, **SD-CPSF** and **EMD-CGSF** we follow analogous procedure as it was applied before to eLORETA solution.

## 3.4 Study of the Type I and II Leakage in simulations

*Figure 4* below shows the results for a typical trial of Simulation 1 and Simulation 2 described in *Subsection 3.1*, as tridimensional colormaps of the Activity within an interval between *0* and maximum value *1*. First, we present the Data Empirical Covariance diagonal values $diag(\mathbf{S}_{vv}^{sim})$ at the Sensors space $\mathbb{E}$ (343 points), see *Figure 4 a) c)*, as portraying of the Volume Conduction effect in the Gold Standard Scalp projection. Second, the estimated **PSF** and **GSF** and the corresponding simulation Gold Standard are presented at the reduction of the Cortical Manifold space $\mathbb{G}$ (6K points), for the Methods eLORETA, LCMV and BC-VARETA, see *Figure 4 b) d)*.

Correspondingly, *Figure 5* below shows the results of the Connectivity for the typical trial of *Simulation 2*, as bidimensional colormaps within an interval between *0* and maximum value *1*. First, we present the Data Empirical Covariance Matrix Inverse $inv(\mathbf{S}_{vv}^{sim})$ at the Cartesian product of Sensor spaces $\mathbb{E} \times \mathbb{E}$ (343×343 points), see *Figure 5 a)*, as portraying the Volume conduction effect in the Connectivity of the Gold Standard Scalp projection. Second, the estimated **CPSF** and **CGSF** and the corresponding simulation Gold Standard are presented at the Cartesian product of reduced Cortical Manifold spaces $\mathbb{G} \times \mathbb{G}$, (6K×6K points), for the Methods eLORETA, LCMV and BC-VARETA, see *Figure 5 b)*. For an easier visualization we don't show the full space, but a subspace defined by 20 neighbors of each Active Source in the actual configuration.

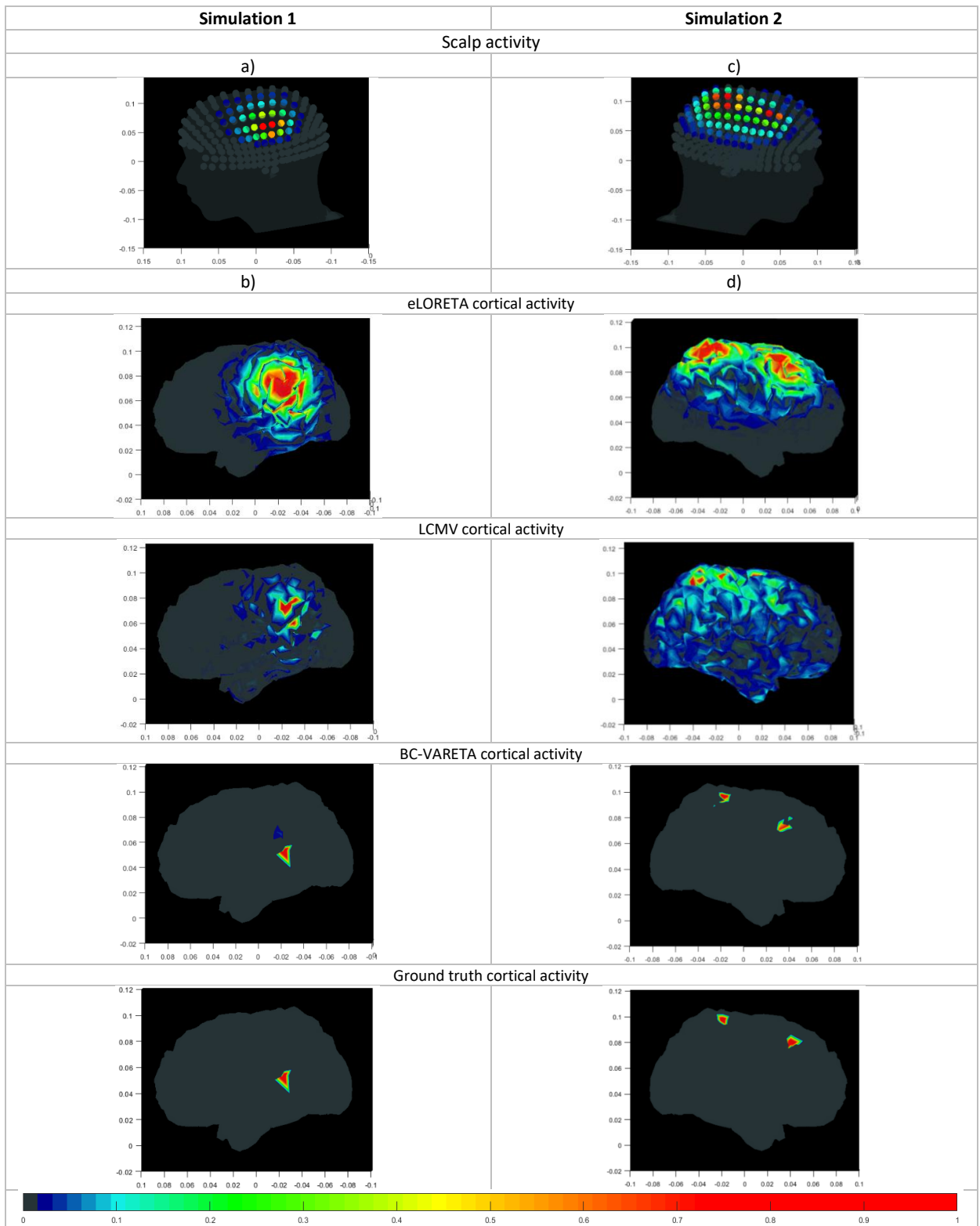

*Figure 4:* A typical trial of the two configurations (one active dipole, two active dipoles) to evaluate the Volume Conduction effect and Localization performance. Projected Scalp Activity given one active generator a), two active generators c). Activity estimated with the methods eLORETA, LCVM and BC-VARETA given one active generator b), two active generators d). The tridimensional colormaps are shown within an interval between 0 and maximum value 1.

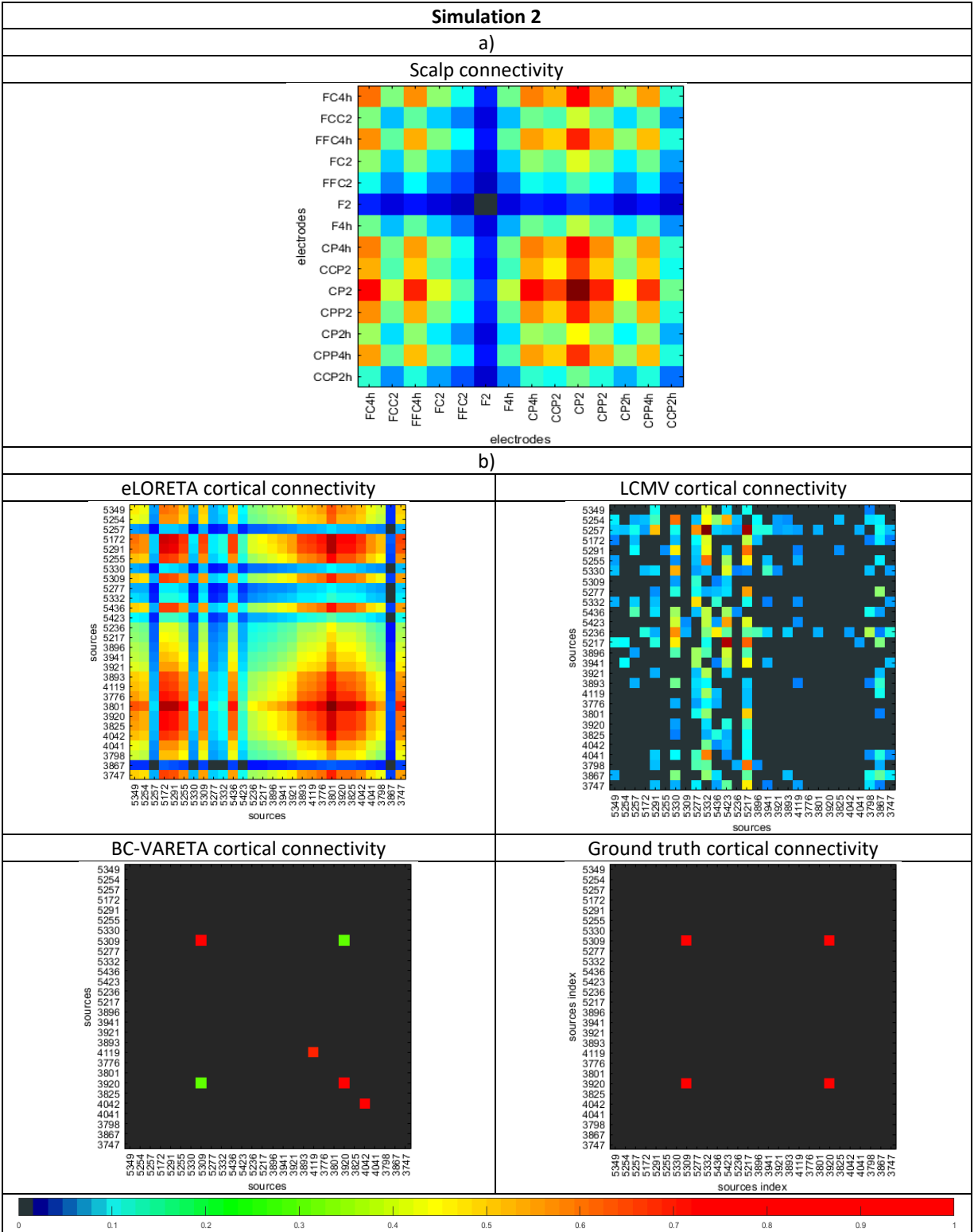

*Figure 5:* A typical trial of the simulated configuration (two active dipoles) to evaluate the Volume Conduction effect and Connectivity performance. Projected Scalp Connectivity given two active generators a), four active generators c). Connectivity estimated with the methods eLORETA, LCVM and BC-VARETA given two active generators b), four active generators d). The bidimensional colormaps are shown within an interval between 0 and maximum value 1.

We performed further analysis in *Simulation 3*, regarding the distance between Active Sources at each of the 500 configurations. This was done stablishing three classifications of distance. **Short Range**: The maximum distance between Sources was smaller than 5 cm. **Middle Range**: The minimum distance between Sources was greater than 5cm and the maximum smaller than 8 cm. **Long Range**: The minimum distance between Sources was greater than 8 cm. We show the results for typical trials corresponding to each classification of distance analogously to *Figure 4* and *Figure 5*. See *Figure 6* for the tridimensional colormaps of Activity and in *Figure 7* for the bidimensional colormaps of the Connectivity.

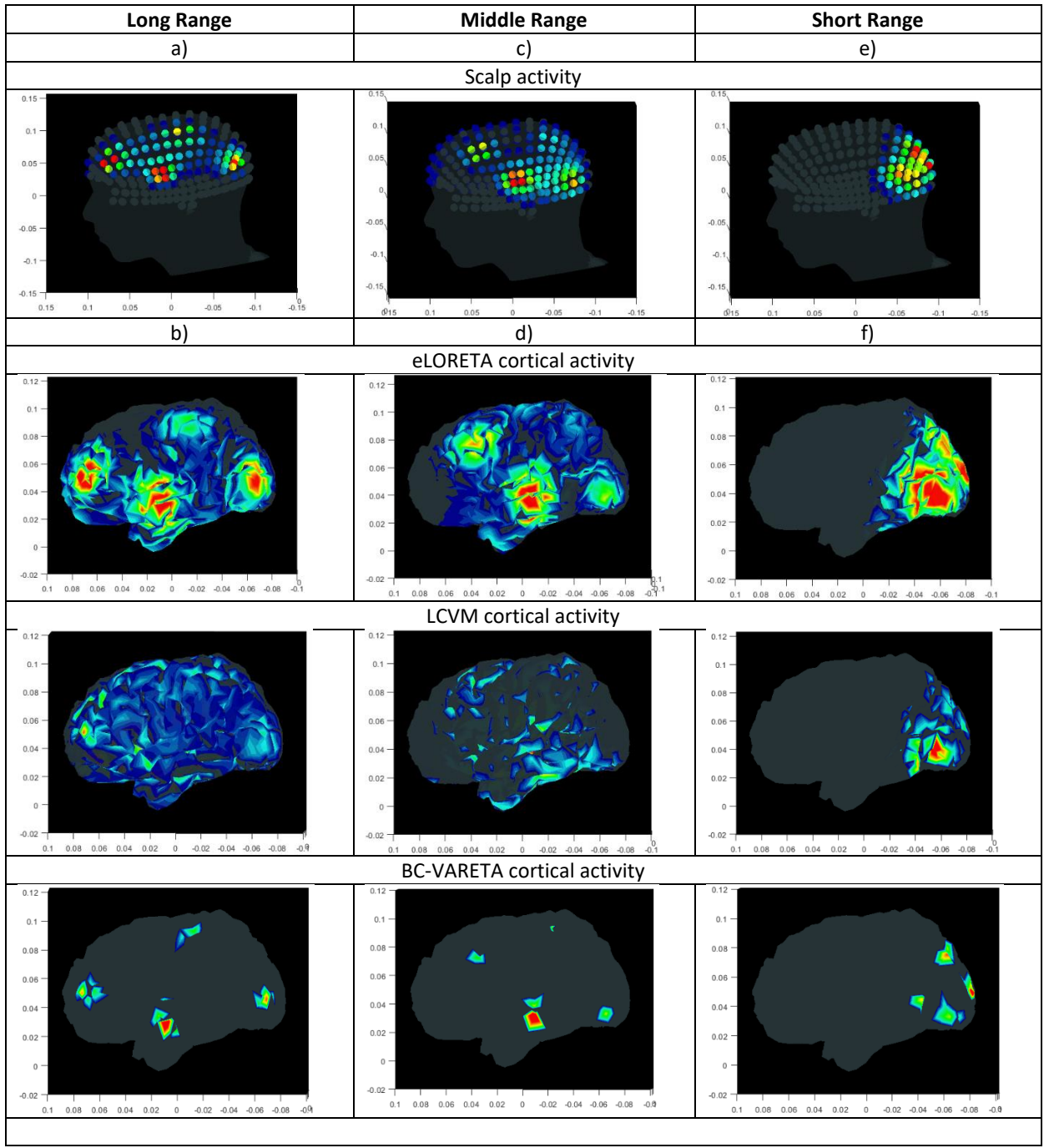

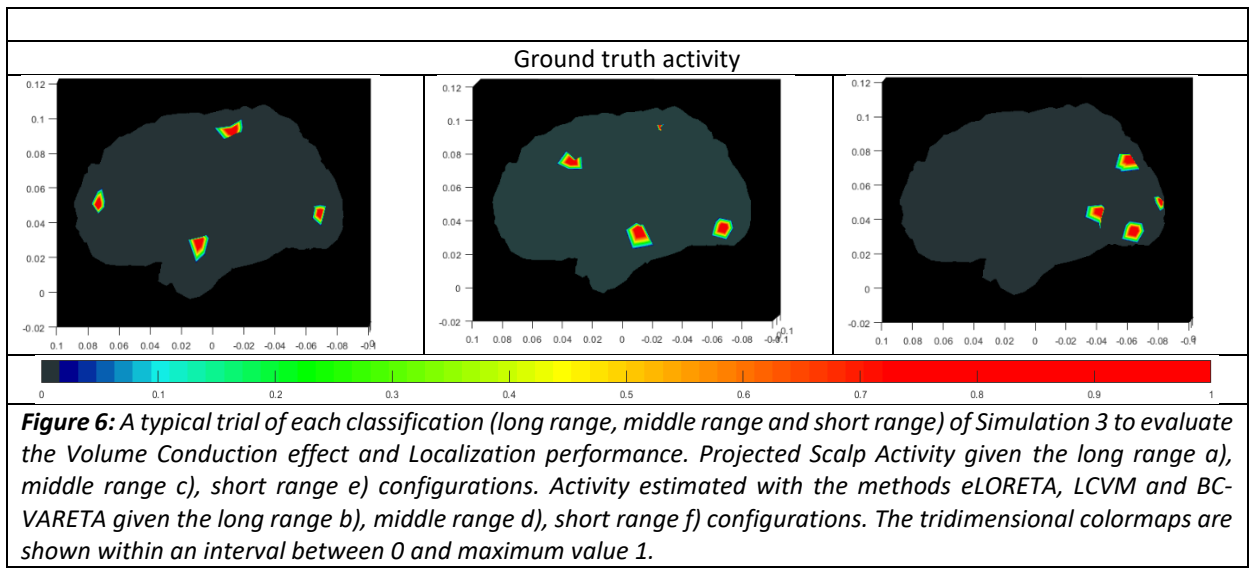

*Figure 6:* *A typical trial of each classification (long range, middle range and short range) of Simulation 3 to evaluate the Volume Conduction effect and Localization performance. Projected Scalp Activity given the long range a), middle range c), short range e) configurations. Activity estimated with the methods eLORETA, LCVM and BC-VARETA given the long range b), middle range d), short range f) configurations. The tridimensional colormaps are shown within an interval between 0 and maximum value 1.*

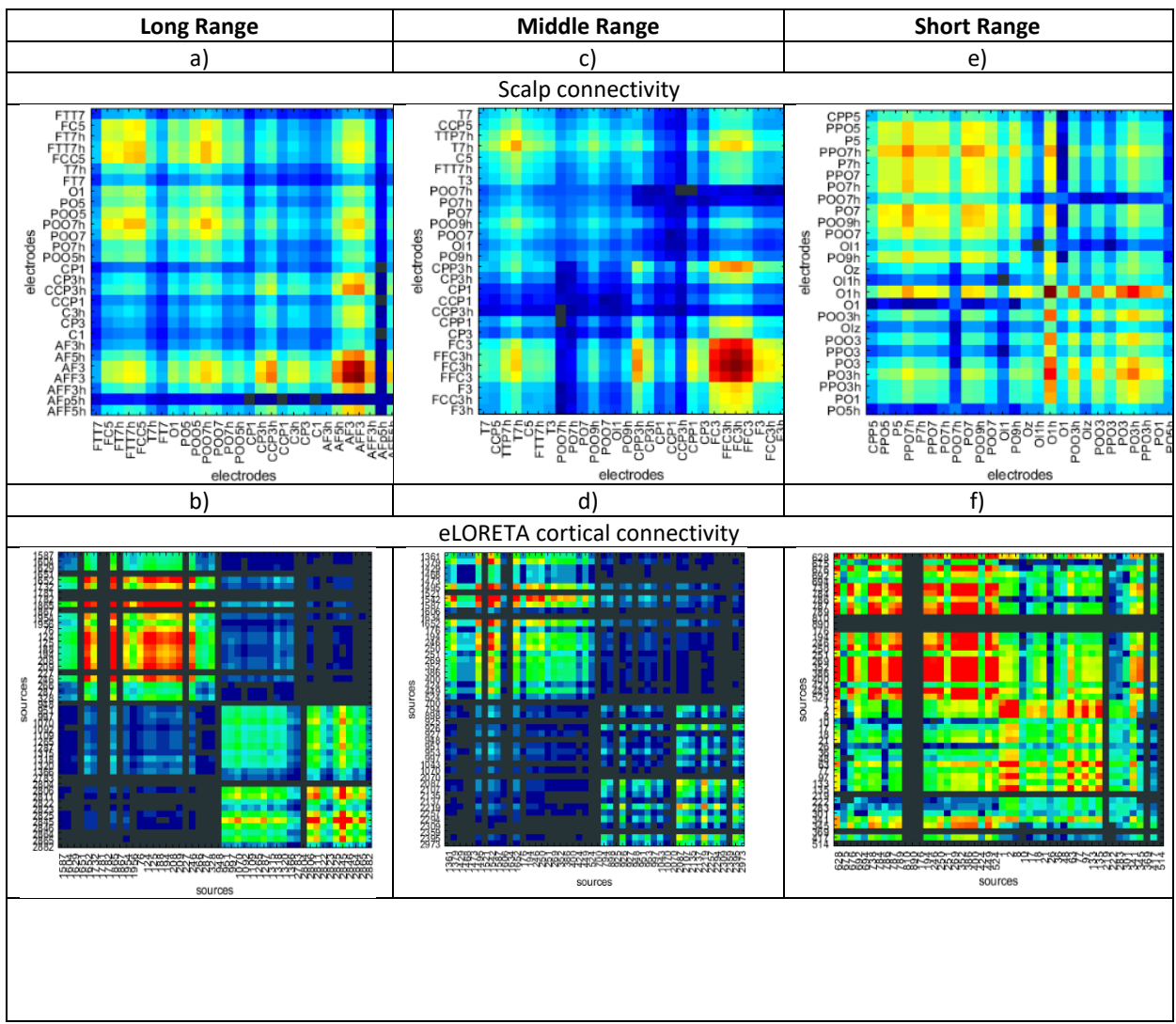

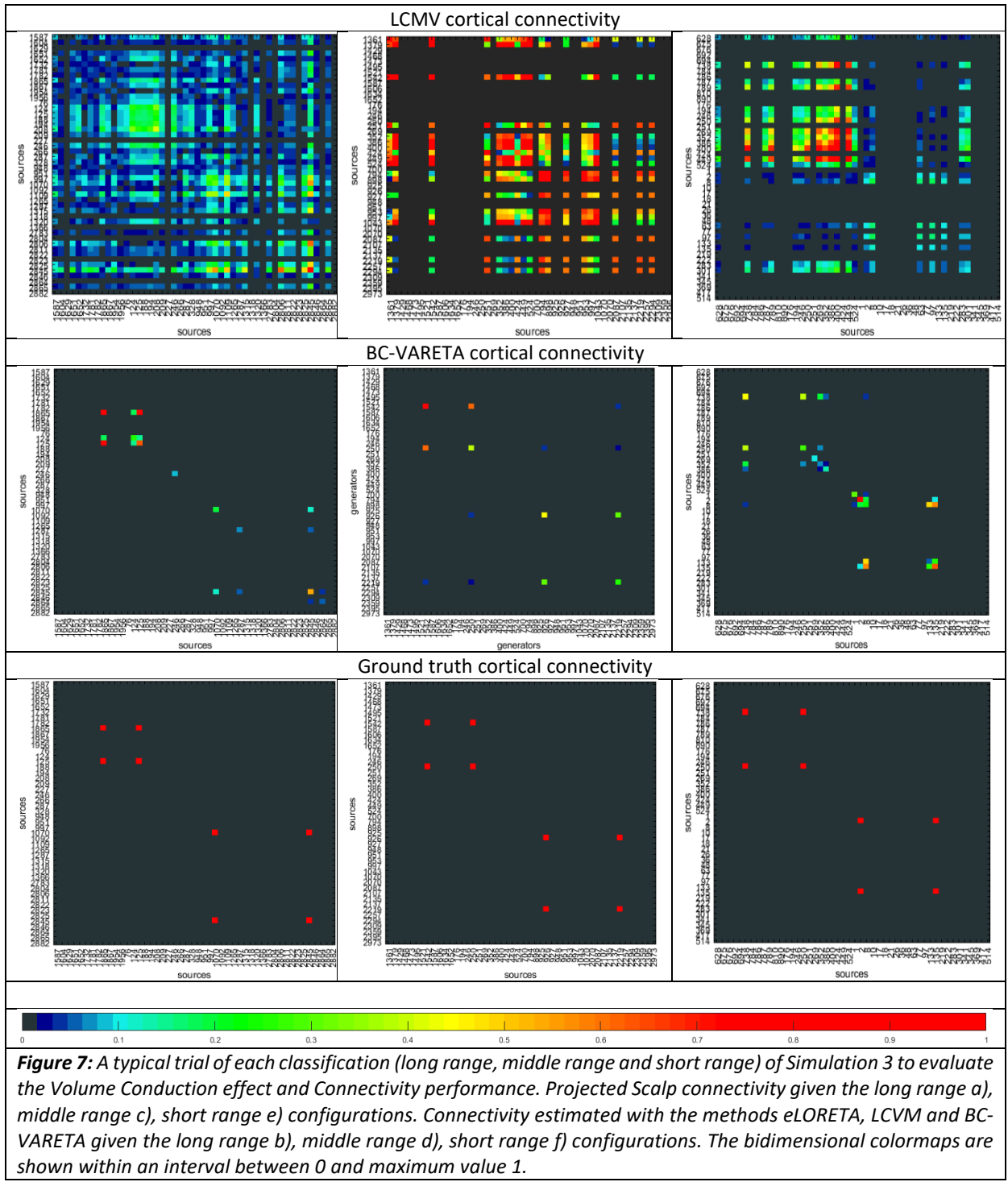

*Figure 7: A typical trial of each classification (long range, middle range and short range) of Simulation 3 to evaluate the Volume Conduction effect and Connectivity performance. Projected Scalp connectivity given the long range a), middle range c), short range e) configurations. Connectivity estimated with the methods eLORETA, LCVM and BC-VARETA given the long range b), middle range d), short range f) configurations. The bidimensional colormaps are shown within an interval between 0 and maximum value 1.*

We report the Mean and Standard Deviation of the **SD-PSF**, **EMD-GSF**, **SD-CPSF** and **EMD-CGSF** measures for the Methods eLORETA, LCMV and BC-VARETA, computed for the 500 configurations of *Simulation 1* and *Simulation 2*, see *Table 5* below. The corresponding results for *Simulation 3* of the **EMD-GSF** and **EMD-CGSF** measures Mean and Standard Deviation are reported separately for each classification (long range, middle range and short range), see *Table 6*.

Table 5: Mean and Standard Deviation of the SD-PSF, EMD-GSF, SD-CPSF and EMD-CGSF measures for the methods eLORETA, LCMV and BC-VARETA in Simulation 1 and 2.

| Methods | Simulation 1 | | Simulation 2 | | |
|---|---|---|---|---|---|
| | *EMD-GSF* | *SD-PSF* | *EMD-GSF* | *SD-CPSF* | *EMD-CGSF* |
| **eLORETA** | 26.9305 ± 3.9276 | 0.1093 ± 0.0166 | 30.4747 ± 6.5958 | 107.8553 ± 25.3005 | 107.8553 ± 25.3005 |
| **LCMV** | 10.2282 ± 11.4917 | 0.1592 ± 0.0447 | 37.1807 ± 10.2112 | 132.7289 ± 32.6940 | 132.7289 ± 32.6940 |
| **BC-VARETA** | 0.4632 ± 0.7121 | 0.0105 ± 0.0128 | 0.9712 ± 0.5589 | 2.8131 ± 0.4494 | 2.8131 ± 0.4494 |

Table 6: Mean and Standard Deviation of the SD-PSF, EMD-GSF, SD-CPSF and EMD-CGSF measures for the methods eLORETA, LCMV and BC-VARETA in each distance classification of Simulation 3.

| Methods | Long Range | | Middle Range | | Short Range | |
|---|---|---|---|---|---|---|
| | *EMD-GSF* | *EMD-CGSF* | *EMD-GSF* | *EMD-CGSF* | *EMD-GSF* | *EMD-CGSF* |
| **eLORETA** | 11.4388 ± 2.9366 | 231.0420 ± 60.7255 | 7.9462 ± 1.5958 | 216.6924 ± 25.3005 | 16.0355 ± 0.2562 | 139.9369 ± 10.1142 |
| **LCMV** | 14.2701 ± 1.8652 | 201.7472 ± 18.8072 | 20.3405 ± 10.2112 | 274.7283 ± 32.6940 | 18.1182 ± 1.9750 | 163.6560 ± 29.1482 |
| **BC-VARETA** | 2.2695 ± 1.7096 | 43.7651 ± 4.5070 | 1.8992 ± 0.6810 | 21.4402 ± 9.8305 | 4.4229 ± 1.5213 | 16.4168 ± 5.6970 |

# 4 Discussion

## 4.1 Methodology of Brain Connectivity Variable Resolution Tomographic Analysis

The State of Art of Electrophysiological Brain Source Localization and Connectivity is quite diverse. Despite this fact, the structure of the proposed Bayesian Model, underlying the BC-VARETA framework, is common for a large family of Methods, see *Subsection 2.1* and *Table 1* in Subsection 2.2. Furthermore, we intended enough generality of our Model set up, within the Square of acceptable Physiological and Mathematical assumptions, by avoiding the use of constraints that could prevent for a fair analysis of the ***Parameters*** and ***Hyperparameters***. It is based on a Hierarchically Conditioned and fully Multivariate Two Levels Gaussian Graphical Model of ***Data*** and ***Parameters***, see Subsection 2.1, in consistency with the Bayesian representation of the LSSM, in both Time and Frequency Domain (Wills et al., 2009; Faes et al., 2012; Galka et al., 2004; Pascual-Marqui et al., 2014; Valdes-Sosa 2004; Lopes da Silva et al., 1980; Baccalá and Sameshima, 2001; Babiloni et al., 2005).

The ad-hoc Conditional *Data* Covariance structure was regarded as an observable property of Instrumental/Environmental/Biological Noisy processes (Waldorp et al., 2001; De Munck et al., 2002; Huizenga et al., 2002). In general, this strategy improves the algorithm convergence by ruling out adverse effects, given the Non-linear interaction between the estimation of multiple ***Hyperparameters*** within the ***Data*** Conditional Covariance, and the ***Parameters*** Covariance, as it happens with LORETA, eLORETA and eLORETA (***Table 1***). In similitude to previous works we regarded the Noise Variance as a ***Hyperparameter***, see VARETA, ARD, ReLM and SSBL in *Table 1*. Distinctively, a Noise Variance Jeffrey Improper Gibbs Prior allowed us to set a Noise inferior limit, that constitutes also an observable property of Noisy processes (Van Hoey et al., 2000; Lemm et al., 2006; Bigdely-Shamlo et al., 2015; Phillips et al., 2002; Phillips et al., 2005). Setting a Noise inferior limit also prevents for adverse effects like assimilating higher amount of

Noise at the Sources Level, which can lead to the overestimation of the **Parameters** Covariance. This also contributes to the stability of the algorithms in general.

We formulated general Sparse Gibbs Priors on the **Parameters** Precision Matrix and not on the Covariance, see *Subsection 2.1* and *Table 2* of *Subsection 2.3*. To regard the Covariance as informative of the Connectivity instead, is a wrong approach in sight of the LSSM theory (Wills et al., 2009; Faes et al., 2012; Galka et al., 2004; Pascual-Marqui et al., 2014; Valdes-Sosa 2004; Lopes da Silva et al., 1980; Baccalá and Sameshima, 2001; Babiloni et al., 2005). This constitutes a common misdeed of previous works that pursue the Connectivity analysis through Sparse Sources Covariance estimation, like ReLM, or Covariance extraction as a Postprocessing of Sources Activity estimates, like MNE, LORETA, eLORETA, eLORETA, ARD, VARETA, SSBL. In addition, our formulation attains to incorporate information on the Gray Matter areas Intra/Inter-connections. The Model Priors were rather general, allowing to include Structured Sparsity given an anatomical segmentation and probability mask of Intra-Cortical connections strength decay with distance and probabilistic maps of the White Matter tracks connectivity.

With the BC-VARETA framework we attained interpretable formulations of the First Level of Inference. The Maximum Posterior analysis (First Level of Inference) leads to an estimator of the Parameters that has identical expression to those of previous methodologies, see *Table 1* in *Subsection 2.2*. Thus, formulating the *Type I* Leakage measures upon BC-VARETA is representative and extendable to the whole State of the Art in Electrophysiological Brain Source Localization. The Second Level of Inference constituted the most particular aspect in setting up the BC-VARETA. We proposed a unification of the State of Art in Electrophysiological Source Localization and Connectivity with the theory GGM's, denominated here as SGGM. This was done by applying the EM strategy (Dempster et al., 1977; Liu and Rubin, 1994; McLachlan and Krishnan, 2007) and expressing the Expected Log-Likelihood as a function of the ESEC Matrix, see *Subsection 2.2.2*. The Posterior analysis with Gibbs Priors on the Precision Matrix **Hyperparameters** lead to Target Function analogous to that of a typical GGM's.

The current GGM's theory however was not able to serve the mathematical semblance of the Electrophysiological SGGM (Friedman et al, 2008; Mazumder et al. 2012; Schmidt, 2010; Hsieh, 2014; Danaher et al., 2014; Drton and Maathuis, 2017), see *Table 2*. In consequence we explored the GGM's from the Bayesian perspective and proposed a new solution that better suits the Electrophysiological SGGM scenario, see *Subsection 2.3*. To achieve this, we used a generalization of Andrews and Mallows Lemma to the Gibbs Priors of the Precision Matrix with Penalization functions in the LASSO family (Andrews and Mallows, 1974; Tipping, 2001; Schmolck and Everson, 2007; Faul and Tipping, 2002; Li and Lin, 2010; Kyung et al., 2010). This generalization allowed for Concave Quadratic reformulation of the SGGM Target Function in resembling the strategy of LQA algorithms (Fan and Li, 2001; Valdés-Sosa et al., 2006; Sánchez-Bornot et al., 2008). In the context of the SGGM LQA we implemented a Standardization technique that simplifies the Target Function minimization problem, given the Scale Invariance properties of the Wishart Likelihood (Srivastava, 1965; Drton et al., 2008). The Connectivity estimator was derived, as consequence of applying the LQA and Standardization to the original SGGM, from the direct solution of a Matrix Riccati equation (Lim, 2006; Honorio and Jaakkola, 2013). To prevent the biasing of the Precision Matrix estimator given the selection of specific SGGM Penalty function and Regularization parameters (Jankova and Van De Geer, 2015, 2017), we use a debiasing operation proposed in (Jankova and Van De Geer, 2018).

The Posterior analysis of the Noise Variance Hyperparameter lead to an interpretable formula composed of the also typical EM Naïve estimator plus the inferior limit (Dempster et al., 1977; Liu and Rubin, 1994; McLachlan and Krishnan, 2007; Valdes-Sosa, 1996, Bosch-Bayard, et al., 2001), in correspondence to what was discussed before about the Noise Variance Jeffrey Improper Gibbs Prior. The bias of this formula lies on quantities that can be interpreted experimentally, thus there is not ambiguity on the definition such estimator.

## 4.2 Theoretical analysis of the Type I and II Leakage effect measures

In this work we provided a unified description of the Leakage by a Model that regards the Source Activity and Connectivity as elements to be estimated into a System Identification approach. The two different scenarios of Leakage were correspondingly represented by the two Levels of Inference of the BC-VARETA methodology. The **PSF** and **GSF** formulated at the Firs Level of Inference, were representative of the *Type I* Leakage effect given Volume Conduction in multiple scenarios as proposed by (Schoffelen and Gross, 2009; Wens et al., 2015; Brunner et al., 2016; Silva Pereira et al., 2017; Hincapié et al., 2017). The original **PSF** concept was too ideal to cover all the aspects of realistic Electrophysiological Sources Localization, thus it was generalized to consider multiple Active cortical sources, the Inverse Crime (Kaipio & Somersalo 2004) and the Noise from biological, environmental and instrumentation origins.

We provide a representation of the *Type II* Leakage or Connectivity Leakage through the **CPSF** and **CGSF** at the Second Level of Inference. This is a more consistent approach in comparison to previous works, that have considered the Connectivity Leakage (*Type II* Leakage) and its Correctors in the limited context of the linear mixing or crosstalk between Sources Activity estimates (*Type I* Leakage) (Brookes et al., 2012; McCoy and Troop, 2013; Colclough et al., 2015; Wens et al., 2015; Colclough et al., 2016; Silva Pereira et al., 2017). Even though, these approaches constituted a first approximation to the Leakage effect, they lacked objectivity when ignored more realistic Non-linear Methods in the State of the Art of Sources Activity and Connectivity, see *Table 1*. Distinctively, the BC-VARETA framework allowed us to formulate Non-linear mutually interacting and explicit estimators of the **PSF** (**GSF**) and **CPSF** (**CGSF**), see *Subsection 2.2.1* and *Subsection 2.4.1*, that evince a bidirectional relationship between the Type I and II Leakage.

Modelling the Source Activity and Connectivity as a whole, as was done with BC-VARETA, provided a consistent way to implement the correction of the Connectivity Leakage, by incorporating the Corrector Operator at the Model Priors. In this sense, a more consistent approach was presented by (Pascual-Marqui et al., 2017; Bosch-Bayard and Biscay, 2018), accounting for the linear de-mixing of LSSM Kalman Filter Non-linear estimators. Distinctively, in our work we revendicate the concept of Sparsity in the Connectivity Level as part of the Leakage correction strategy, that might coexist and never collide with different approaches in the context of Non-linear Methods (like linear de-mixing LSSM Kalman Filter estimators).

A measure of the *Type I* Leakage in the **PSF** was built on the Geodesic Distance Spatial Dispersion (**SD-PSF**) in the Cortical Manifold, that has been universally adopted for single point spreading like scenarios. We provide an extension of this concept to multiple points by the **GSF** and its **EMD** (**EMD-GSF**). Remarkably, we present a generalization to the Cartesian geometry (Product of Cortical Manifolds Spaces) to represent the measures Type II Leakage given by **CPSF** and **CGSF**. To this end we use the Spatial Dispersion of the Cartesian Geodesic Distance (**SD-CPSF**), for a single connection, and the Cartesian Earth Mover's distance (**EMD-CGSF**), for the extension to multiple connections.

## 4.3 Analysis of the Results in Simulations

*Simulation Aims*

*Simulation 1* was set up to study the Type I Leakage in an ideal scenario that reflects solely the Volume Conduction spilling effect on a single point. *Simulation 2* was aimed to study the Type II Leakage and its mutual interaction with the Type I Leakage, in a scenario that reflects the spilling effect of Volume Conduction on two points Connectivity. *Simulation 3* pursues the study the Type II Leakage on four points where only three of them were connected, reflecting not only the Volume Conduction spilling effect in Connectivity but also the crosstalk towards not connected points.

*Scalp Analysis*

For a typical trial of *Simulation 1* the projected Activity at the Scalp Sensors of a typical trial, see *Figure 4 a)*, showed the single point large spatial spillover and mismatch of its maximum given to the Volume Conduction effect, confirming an essential shortcoming of the direct analysis of Sensors data. This has been pointed out in previous works (Brunner et al., 2016; Van de Steen et al., 2016). Consistently to *Simulation 1* results, the Scalp projections of typical trials in *Simulation 2*, see *Figure 4 c)*, showed an even larger spatial spillover and mismatch their maximum given to the Volume Conduction effect. The situation of *Simulation 3*, see *Figure 6 a) c) e)*, was more anfractuous given the mixture of the projected Sources in the conditions of short and middle range distance, thus making impossible deduce the existence of the third or fourth source in the simulation, just by analyzing its Scalp projection. This happened in lower degree in the long range condition.

Consequently, this effect is reflected in the Precision Matrix (Connectivity) associated to the Data Empirical Covariance Matrix at the Cartesian Scalp Sensors space product, see *Figure 5 a) and Figure 7 a) c) e)*. The Scalp Connectivity analysis based on Precision Matrices (Kaminski and Blinowska, 2014; Kaminski and Blinowska, 2017; Blinowska, 2011), seems unfavorable as revealed by the visual inspection of these bidimensional maps: First: A large spillover of the Non-diagonal Block in the two points Connectivity of *Simulation 2*, see *Figure 5 a)*. Second: Mixture of the Diagonal Blocks corresponding to the individual Sources Precisions in the three conditions (short, middle and long range distance) of *Simulation 3*, see *Figure 7 a) c) e)*. Third: Crosstalk towards not connected points given the large spillover of the Non-diagonal Blocks in the three conditions (short, middle and long range distance) of *Simulation 3*, see *Figure 7 a) c) e)*. Even when we are in presence of highly sparse simulations the scenario for Scalp Connectivity doesn't show any goodness according to these results.

*Sources Analysis*

The Sources Localization and Connectivity were distorted qualitatively by the Volume Conduction effect, across all simulations. For the typical trial of *Simulation 1* the reconstruction with eLORETA does not improves the mentioned situation for the Scalp Data, considering that the Sources were extended along a larger Cortical area than the simulated Scalp projection, see *Figure 4 b)*. This overestimation of Cortical activity is a peculiar of Linear Methods such as MNE, see *Table 1*, but in this case the eLORETA showed qualitatively similar performance despite its Hybrid estimation formulas. We found that the Cortical extension of the LCMV reconstruction was much shorter. This was possible due to the Linear Constraints of the LCMV Hybrid formulas at the Second Level of Inference, that pursue sparsity of the Spatial Filtering Variances. The BC-VARETA reconstruction was the sparsest confirming our hypothesis, about the effect of using a Sparse Precision Matrix Model, that underlies this Non-linear Sources estimation Method.

The Sources Activity reconstruction in typical trials of *Simulation 2* see *Figure 4 d)* and *Simulation 3*, see *Figure 5 b) d) f)*, for eLORETA was extended across a large Cortical area as expected, and consistently with the results in *Simulation 1* it spills beyond the Scalp projection. The Connectivity, see *Figure 5 b)* and *Figure 7 b) d) f)*, presented also high spatial spillover, mixing and crosstalk between Sources. This is the cause of repetitive allusions within the State of the Art to the *Type I* and *II* Leakage (Brookes et al., 2012; McCoy and Troop, 2013; Colclough et al., 2015; Wens et al., 2015; Colclough et al., 2016; Silva Pereira et al., 2017). The LCMV method achieves better performance in Source Activity and Connectivity reconstruction. The mixing of Sources and Crosstalk is qualitatively diminished, but still too spilled as compared to the highly sparse simulated Activity. The Source Activity and Connectivity reconstruction with the BC-VARETA Method was the sparsest and thus the best according to the properties of the simulations, also evidencing that the estimation with sparse Precision Matrix model worked as expected. The spatial spillover, mixing and crosstalk of reconstructed Sources Activity and Connectivity, of the presented typical trials, seems minimized across all simulations by this Non-linear Sparse Method. This outcome was effective for the three conditions (short, middle and long range distance) of *Simulation 3*, whereas expected the results with all Methods deteriorated with the range shrinking.

The results across the 500 trials of the measures **SD-PSF**, **EMD-GSF**, **SD-CPSF** and **EMD-CGSF**, for *Simulation 1* and *Simulation 2* and *Simulation 3* were consistent with the qualitative analysis given for the typical ones, see *Table 5* and *Table 6*. The measures values were always minimum for the BC-VARETA Method. The measures of Sources Localization **SD-PSF** and **EMD-GSF** were consistent to the **SD-CPSF** and **EMD-CGSF**, in the sense of that when the Connectivity reconstruction performance was higher it was always higher the performance in Source Localization. The values of the **EMD-GSF** and **EMD-CGSF**, along the three conditions (short, middle and long range distance) of *Simulation 3* revealed that reconstruction performance deteriorated for all methods as the **Range** decreased, see *Figure 6*. Even though this behavior was expected for every Source Localization Method, the BC-VARETA exhibited the most robust performance.

## 5 Conclusions

The proposed methodology BC-VARETA allowed us to caulk the "Leakage Effect" in simulation scenario of MEEG activity that was challenging, according to the high degree of sparsity (super resolution) variability (different configurations with random positions of sources) and realism (presence of noise in generators and sensors, and inverse crime evaluation). The BC-VARETA performance was better than well stablished methods, which operate under different assumptions, i.e. eLORETA and LCMV. These results were supported by sensitive quality measures (Spatial Dispersion and Earth Movers Distance), that are also acknowledged to be the most interpretable into the state of the art of MEEG source connectivity analysis. Remarkably, our Bayesian model and inference (BC-VARETA) constitutes a unification of the state of the art in the theory of MEEG source activity and connectivity estimation methods and the theory of Gaussian Graphical Models. We presented fully detailed technical derivations of BC-VARETA, along with its interpretability and theoretical comparison with those methodologies previously developed. Another issue addressed into this paper was the rigorous mathematical representation of the Leakage in both source activity and connectivity. It involved the introduction of new quantities as such the Generalized Spread Function (explicit activity estimator given multiple sources) and Cartesian Generalized Spread Function (explicit connectivity estimator given multiple connected sources), and the generalization of Spatial Dispersion and Earth Movers distance to the connectivity space, i.e. Cartesian product of Cortical Manifolds.

## Acknowledgements
The Grant No. 61673090 from the National Nature Science Foundation of China funded this study.

## References

[1] Andrews, D.F. and Mallows, C.L., 1974. Scale mixtures of normal distributions. Journal of the Royal Statistical Society. Series B (Methodological), pp.99-102.
https://www.jstor.org/stable/2984774

[2] Asadi, N. B., Rish, I., Scheinberg, K., Kanevsky, D., and Ramabhadran, B., 2009. Map approach to learning sparse Gaussian Markov networks. In Acoustics, Speech and Signal Processing, 2009. ICASSP 2009. IEEE International Conference on, pp.1721-1724.
https://doi.org/10.1109/ICASSP.2009.4959935

[3] Attias, H., 2000. A variational bayesian framework for graphical models. In Advances in neural information processing systems (pp. 209-215).
https://doi.org/10.1109/ICASSP.2009.4959935 ononon

[4] Babiloni, F., Cincotti, F., Babiloni, C., Carducci, F., Mattia, D., Astolfi, L., Basilisco, A., Rossini, P.M., Ding, L., Ni, Y. and Cheng, J., 2005. Estimation of the cortical functional connectivity with the multimodal integration of high-resolution EEG and fMRI data by directed transfer function. Neuroimage, 24(1), pp.118-131.
https://doi.org/10.1016/j.neuroimage.2004.09.036

[5] Babacan, S.D., Luessi, M., Molina, R. and Katsaggelos, A.K., 2012. Sparse Bayesian methods for low-rank matrix estimation. IEEE Transactions on Signal Processing, 60(8), pp.3964-3977.
https://doi.org/10.1109/TSP.2012.2197748

[6] Baccalá, L.A. and Sameshima, K., 2001. Partial directed coherence: a new concept in neural structure determination. Biological cybernetics, 84(6), pp.463-474.
https://doi.org/10.1007/PL000079

[7] Balkan, O., Kreutz-Delgado, K. and Makeig, S., 2014. Localization of more sources than sensors via jointly-sparse Bayesian learning. IEEE Signal Processing Letters, 21(2), pp.131-134.
https://doi.org/10.1109/LSP.2013.2294862

[8] Belardinelli, P., Ortiz, E., Barnes, G., Noppeney, U. and Preissl, H., 2012. Source reconstruction accuracy of MEG and EEG Bayesian inversion approaches. PloS one, 7(12), p.e51985.
https://doi.org/10.1371/journal.pone.0051985

[9] Bigdely-Shamlo, N., Mullen, T., Kothe, C., Su, K.M. and Robbins, K.A., 2015. The PREP pipeline: standardized preprocessing for large-scale EEG analysis. Frontiers in neuroinformatics, 9, p.16.
https://doi.org/10.3389/fninf.2015.00016

[10] Blinowska, K.J., 2011. Review of the methods of determination of directed connectivity from multichannel data. Medical & biological engineering & computing, 49(5), pp.521-529.
https://doi.org/10.1007/s11517-011-0739-x

[11] Bosch-Bayard, J., Valdes-Sosa, P., Virues-Alba, T., Aubert-Vazquez, E., John, E.R., Harmony, T., Riera-Diaz, J. and Trujillo-Barreto, N., 2001. 3D statistical parametric mapping of EEG source spectra by means of variable resolution electromagnetic tomography (VARETA). Clinical Electroencephalography, 32(2), pp.47-61.
https://doi.org/10.1177/155005940103200203

[12] Bosch-Bayard, J. and Viscay, R.J., 2018. Several drawbacks in brain connectivity assessment based on source estimates -elucidations, practical guidelines, and incentives for further research. Submitted to Brain Topography.

[13] Brookes, M.J., Woolrich, M.W. and Barnes, G.R., 2012. Measuring functional connectivity in MEG: a multivariate approach insensitive to linear source leakage. Neuroimage, 63(2), pp.910-920.
https://doi.org/10.1016/j.neuroimage.2012.03.048

[14] Brunner, C., Billinger, M., Seeber, M., Mullen, T.R. and Makeig, S., 2016. Volume conduction influences scalp-based connectivity estimates. Frontiers in computational neuroscience, 10, p.121.
https://doi.org/10.3389/fncom.2016.00121

[15] Colclough, G.L., Brookes, M.J., Smith, S.M. and Woolrich, M.W., 2015. A symmetric multivariate leakage correction for MEG connectomes. NeuroImage, 117, pp.439-448.
https://doi.org/10.1016/j.neuroimage.2015.03.071

[16] Colclough, G.L., Woolrich, M.W., Tewarie, P.K., Brookes, M.J., Quinn, A.J. and Smith, S.M., 2016. How reliable are MEG resting-state connectivity metrics?. Neuroimage, 138, pp.284-293.
https://doi.org/10.1016/j.neuroimage.2016.05.070

[17] Danaher, P., Wang, P. and Witten, D.M., 2014. The joint graphical lasso for inverse covariance estimation across multiple classes. Journal of the Royal Statistical Society: Series B (Statistical Methodology), 76(2), pp.373-397.
https://doi.org/10.1111/rssb.12033

[18] Daunizeau, J. and Friston, K.J., 2007. A mesostate-space model for EEG and MEG. NeuroImage, 38(1), pp.67-81.
https://doi.org/10.1016/j.neuroimage.2007.06.034

[19] Da Silva, F.L., Vos, J.E., Mooibroek, J. and Van Rotterdam, A., 1980. Relative contributions of intracortical and thalamo-cortical processes in the generation of alpha rhythms, revealed by partial coherence analysis. Electroencephalography and clinical neurophysiology, 50(5-6), pp.449-456.
https://doi.org/10.1016/0013-4694(80)90011-5

[20] Dempster, A.P., Laird, N.M. and Rubin, D.B., 1977. Maximum likelihood from incomplete data via the EM algorithm. Journal of the royal statistical society. Series B (methodological), pp.1-38.
https://www.jstor.org/stable/2984875

[21] De Munck, J.C., Huizenga, H.M., Waldorp, L.J. and Heethaar, R.A., 2002. Estimating stationary dipoles from MEG/EEG data contaminated with spatially and temporally correlated background noise. IEEE Transactions on Signal Processing, 50(7), pp.1565-1572.
https://doi.org/10.1109/TSP.2002.1011197

[22] Drton, M., Massam, H. and Olkin, I., 2008. Moments of minors of Wishart matrices. The Annals of Statistics, 36(5), pp.2261-2283.
https://doi.org/10.1214/07-AOS522

[23] Elanbari, M., Rawi, R., Ceccarelli, M., Bouhali, O. and Bensmail, H., 2015. Advanced Computation of a Sparse Precision Matrix HADAP: A Hadamard-Dantzig Estimation of a Sparse Precision Matrix. In Proceedings of the Sixth International Conference on Computational Logics, Algebras, Programming, Tools, and Benchmarking.

[24] Faes, L., Erla, S. and Nollo, G., 2012. Measuring connectivity in linear multivariate processes: definitions, interpretation, and practical analysis. Computational and mathematical methods in medicine, 2012.
http://dx.doi.org/10.1155/2012/140513

[25] Fan J and Li R.: Variable Selection via Nonconcave Penalized Likelihood and Its Oracle Properties. Journal of the American Statistical Association. 2001; 96: pp1348-1360.
https://doi.org/10.1198/016214501753382273

[26] Farahibozorg, S.R., Henson, R.N., and Hauk, O., 2018. Adaptive cortical parcellations for source reconstructed EEG/MEG connectomes. NeuroImage, 169, pp.23-45.
https://doi.org/10.1016/j.neuroimage.2017.09.009

[27] Freeman, W.J., 1980. Use of spatial deconvolution to compensate for distortion of EEG by volume conduction. IEEE Transactions on Biomedical Engineering, (8), pp.421-429.
https://doi.org/10.1109/TBME.1980.326750

[28] Friedman, J., Hastie, T. and Tibshirani, R., 2008. Sparse inverse covariance estimation with the graphical lasso. Biostatistics, 9(3), pp.432-441.

[29] Faul, A.C. and Tipping, M.E., 2002. Analysis of sparse Bayesian learning. In Advances in neural information processing systems (pp. 383-389).

[30] Friston, K., Harrison, L., Daunizeau, J., Kiebel, S., Phillips, C., Trujillo-Barreto, N., Henson, R., Flandin, G. and Mattout, J., 2008. Multiple sparse priors for the M/EEG inverse problem. NeuroImage, 39(3), pp.1104-1120.
https://doi.org/10.1016/j.neuroimage.2007.09.048

[31] Fuchs, M., Kastner, J., Wagner, M., Hawes, S. and Ebersole, J.S., 2002. A standardized boundary element method volume conductor model. Clinical Neurophysiology, 113(5), pp.702-712.
https://doi.org/10.1016/S1388-2457(02)00030-5

[32] Galka, A., Yamashita, O., Ozaki, T., Biscay, R. and Valdés-Sosa, P., 2004. A solution to the dynamical inverse problem of EEG generation using spatiotemporal Kalman filtering. NeuroImage, 23(2), pp.435-453.
https://doi.org/10.1016/j.neuroimage.2004.02.022

[33] Hämäläinen, M.S. and Ilmoniemi, R.J., 1994. Interpreting magnetic fields of the brain: minimum norm estimates. Medical & biological engineering & computing, 32(1), pp.35-42.
https://doi.org/10.1007/BF025124

[34] Hedrich, T., Pellegrino, G., Kobayashi, E., Lina, J. M., and Grova, C., 2017. Comparison of the spatial resolution of source imaging techniques in high-density EEG and MEG. NeuroImage, 157, pp.531-544.
https://doi.org/10.1016/j.neuroimage.2017.06.022

[35] Hincapié, A.S., Kujala, J., Mattout, J., Pascarella, A., Daligault, S., Delpuech, C., Mery, D., Cosmelli, D. and Jerbi, K., 2017. The impact of MEG source reconstruction method on source-space connectivity estimation: a comparison between minimum-norm solution and beamforming. NeuroImage, 156, pp.29-42.
https://doi.org/10.1016/j.neuroimage.2017.04.038

[36] Honorio, J. and Jaakkola, T.S., 2013. Inverse covariance estimation for high-dimensional data in linear time and space: Spectral methods for riccati and sparse models. arXiv preprint arXiv:1309.6838.
https://arxiv.org/abs/1309.6838

[37] Huizenga, H.M., De Munck, J.C., Waldorp, L.J. and Grasman, R.P., 2002. Spatiotemporal EEG/MEG source analysis based on a parametric noise covariance model. IEEE Transactions on Biomedical Engineering, 49(6), pp.533-539.
https://doi.org/10.1109/TBME.2002.1001967

[38] Jankova, J. and Van De Geer, S., 2015. Confidence intervals for high-dimensional inverse covariance estimation. Electronic Journal of Statistics, 9(1), pp.1205-1229.
https://doi.org/10.1214/15-EJS1031

[39] Janková, J. and van de Geer, S., 2017. Honest confidence regions and optimality in high-dimensional precision matrix estimation. Test, 26(1), pp.143-162.
https://doi.org/10.1007/s11749-016-0503-5

[40] Jankova, J. and van de Geer, S., 2018. Inference in high-dimensional graphical models. arXiv preprint arXiv:1801.08512.
https://arxiv.org/abs/1801.08512

[41] Jordan, M.I. ed., 1998. Learning in graphical models (Vol. 89). Springer Science & Business Media.
https://doi.org/10.1007/978-94-011-5014-9

[42] Kaminski, M. and Blinowska, K.J., 2014. Directed transfer function is not influenced by volume conduction—inexpedient pre-processing should be avoided. Frontiers in computational neuroscience, 8, p.61.
https://doi.org/10.3389/fncom.2014.00061

[43] Kaminski, M. and Blinowska, K.J., 2014. Directed transfer function is not influenced by volume conduction—inexpedient pre-processing should be avoided. Frontiers in computational neuroscience, 8, p.61.
https://doi.org/10.3389/fncom.2014.00061

[44] Krishnaswamy, P., Obregon-Henao, G., Ahveninen, J., Khan, S., Babadi, B., Iglesias, J.E., Hämäläinen, M.S. and Purdon, P.L., 2017. Sparsity enables estimation of both subcortical and cortical activity from MEG and EEG. Proceedings of the National Academy of Sciences, p.201705414.
https://doi.org/10.1073/pnas.1705414114

[45] Kyung, M., Gill, J., Ghosh, M. and Casella, G., 2010. Penalized regression, standard errors, and Bayesian lassos. Bayesian Analysis, 5(2), pp.369-411.
https://doi.org/10.1214/10-BA607

[46] Lemm, S., Curio, G., Hlushchuk, Y. and Muller, K.R., 2006. Enhancing the signal-to-noise ratio of ICA-based extracted ERPs. IEEE Transactions on Biomedical Engineering, 53(4), pp.601-607.
https://doi.org/10.1109/TBME.2006.870258

[47] Li, Q. and Lin, N., 2010. The Bayesian elastic net. Bayesian Analysis, 5(1), pp.151-170.
https://doi.org/10.1214/10-BA506

[48] Lim, Y., 2006. The matrix golden mean and its applications to Riccati matrix equations. SIAM Journal on Matrix Analysis and Applications, 29(1), pp.54-66.
https://doi.org/10.1137/050645026

[49] Liu, C. and Rubin, D.B., 1994. The ECME algorithm: a simple extension of EM and ECM with faster monotone convergence. Biometrika, 81(4), pp.633-648.
https://doi.org/10.1093/biomet/81.4.633

[50] MacKay D J C. Information Theory, Inference, and Learning Algorithms. Cambridge University Press 2003.
https://doi.org/10.2277/0521642981

[51] Mazumder, R. and Hastie, T., 2012. The graphical lasso: New insights and alternatives. Electronic journal of statistics, 6, p.2125.
https://doi.org/10.1214/12-EJS740

[52] McLachlan, G. and Krishnan, T., 2007. The EM algorithm and extensions (Vol. 382). John Wiley & Sons.

[53] McCoy, M.B. and Tropp, J.A., 2013. The achievable performance of convex demixing. arXiv preprint arXiv:1309.7478.
https://arxiv.org/abs/1309.7478v1

[54] Neal, R.M., 1998. Assessing relevance determination methods using DELVE. Nato Asi Series F Computer And Systems Sciences, 168, pp.97-132.

[55] Palva, J.M., Wang, S.H., Palva, S., Zhigalov, A., Monto, S., Brookes, M.J., Schoffelen, J.M. and Jerbi, K., 2018. Ghost interactions in MEG/EEG source space: A note of caution on inter-areal coupling measures. Neuroimage, 173, pp.632-643.
https://doi.org/10.1016/j.neuroimage.2018.02.032

[56] Pascual-Marqui, R.D., Michel, C.M. and Lehmann, D., 1994. Low resolution electromagnetic tomography: a new method for localizing electrical activity in the brain. International Journal of psychophysiology, 18(1), pp.49-65.
https://doi.org/10.1016/0167-8760(84)90014-X

[57] Pascual-Marqui, R.D., Esslen, M., Kochi, K. and Lehmann, D., 2002. Functional imaging with low-resolution brain electromagnetic tomography (LORETA): a review. Methods and findings in experimental and clinical pharmacology, 24(Suppl C), pp.91-95.

[58] Pascual-Marqui, R.D., 2002. Standardized low-resolution brain electromagnetic tomography (sLORETA): technical details. Methods Find Exp Clin Pharmacol, 24(Suppl D), pp.5-12.

[59] Pascual-Marqui, R.D., Pascual-Montano, A.D., Lehmann, D., Kochi, K., Esslen, M., Jancke, L., Anderer, P., Saletu, B., Tanaka, H., Hirata, K. and John, E.R., 2006. Exact low resolution brain electromagnetic tomography (eLORETA). Neuroimage, 31(Suppl 1).

[60] Pascual-Marqui, R.D., Biscay, R.J., Bosch-Bayard, J., Faber, P., Kinoshita, T., Kochi, K., Milz, P., Nishida, K. and Yoshimura, M., 2017. Innovations orthogonalization: a solution to the major pitfalls of EEG/MEG" leakage correction". arXiv preprint arXiv:1708.05931.
https://doi.org/10.1101/178657

[61] Paz-Linares, D., Gonzalez-Moreira, E., Valdes-Sosa, P.A., 2018. A Technical Note on the Estimation of Embedded Hermitian Gaussian Graphical Models for MEEG Source Activity and Connectivity Analysis. arXiv preprint arXiv:submit/2413461.
https://arxiv.org/submit/2413461

[62] Phillips, C., Rugg, M.D. and Friston, K.J., 2002. Systematic regularization of linear inverse solutions of the EEG source localization problem. NeuroImage, 17(1), pp.287-301.
https://doi.org/10.1006/nimg.2002.1175

[63] Phillips, C., Mattout, J., Rugg, M.D., Maquet, P. and Friston, K.J., 2005. An empirical Bayesian solution to the source reconstruction problem in EEG. NeuroImage, 24(4), pp.997-1011.
https://doi.org/10.1016/j.neuroimage.2004.10.030

[64] Riera J J and Fuentes M E. Electric lead field for a piecewise homogeneous volume conductor model of the head. Biomedical Engineering, IEEE Transactions on 1998; 45(6): 746-753.
https://doi.org/10.1109/10.678609

[65] Rish, I., and Grabarnik, G., 2014. Sparse modeling: theory, algorithms, and applications. CRC press.

[66] Sánchez-Bornot J M, Martínez-Montes E, Lage-Castellanos A, Vega-Hernández M and Valdés-Sosa P A. Uncovering sparse brain effective connectivity: A voxel-based approach using penalized regression. Statistica Sinica 2008; 18: 1501-1518.
https://www.jstor.org/stable/24308566

[67] Sato, M.A., Yoshioka, T., Kajihara, S., Toyama, K., Goda, N., Doya, K. and Kawato, M., 2004. Hierarchical Bayesian estimation for MEG inverse problem. NeuroImage, 23(3), pp.806-826.
https://doi.org/10.1016/j.neuroimage.2004.06.037

[68] Silva Pereira, S., Hindriks, R., Mühlberg, S., Maris, E., van Ede, F., Griffa, A., Hagmann, P. and Deco, G., 2017. Effect of Field Spread on Resting-State Magneto Encephalography Functional Network Analysis: A Computational Modeling Study. Brain connectivity, 7(9), pp.541-557.
https://doi.org/10.1089/brain.2017.0525

[69] Schmidt, M., 2010. Graphical model structure learning with l1-regularization. University of British Columbia.

[70] Schoffelen, J.M. and Gross, J., 2009. Source connectivity analysis with MEG and EEG. Human brain mapping, 30(6), pp.1857-1865.
https://doi.org/10.1002/hbm.20745

[71] Schmolck, A. and Everson, R., 2007. Smooth relevance vector machine: a smoothness prior extension of the RVM. Machine Learning, 68(2), pp.107-135.
https://doi.org/10.1007/s10994-007-5012-z

[72] Srivastava, M.S., 1965. On the complex Wishart distribution. The Annals of mathematical statistics, 36(1), pp.313-315.
https://www.jstor.org/stable/2238098

[73] Tipping, M.E., 2001. Sparse Bayesian learning and the relevance vector machine. Journal of machine learning research, 1(Jun), pp.211-244.
https://doi.org/10.1162/15324430152748236

[74] Valdés-Hernández, P.A., von Ellenrieder, N., Ojeda-Gonzalez, A., Kochen, S., Alemán-Gómez, Y., Muravchik, C. and Valdés-Sosa, P.A., 2009. Approximate average head models for EEG source imaging. Journal of neuroscience methods, 185(1), pp.125-132. Valdes-Sosa, P.A., 2004. Spatio-temporal autoregressive models defined over brain manifolds. Neuroinformatics, 2(2), pp.239-250.
https://doi.org/10.1016/j.jneumeth.2009.09.005

[75] Valdes-Sosa, P., Marti, F., Garcia, F. and Casanova, R., 2000. Variable resolution electric-magnetic tomography. In Biomag 96 (pp. 373-376). Springer, New York, NY.
https://doi.org/10.1007/978-1-4612-1260-7_91

[76] Valdés-Sosa, P.A., Bornot-Sánchez, J.M., Vega-Hernández, M., Melie-García, L., Lage-Castellanos, A. and Canales-Rodríguez, E., 2006. 18 granger causality on spatial manifolds: applications to neuroimaging. Handbook of time series analysis: recent theoretical developments and applications, pp.461-491.
https://doi.org/10.1002/9783527609970.ch18

[77] Van de Steen, F., Faes, L., Karahan, E., Songsiri, J., Valdes-Sosa, P.A. and Marinazzo, D., 2016. Critical comments on EEG sensor space dynamical connectivity analysis. Brain topography, pp.1-12.
https://doi.org/10.1007/s10548-016-0538-7

[78] Van Hoey, G., Vanrumste, B., D'have, M., Van de Walle, R., Lemahieu, I. and Boon, P., 2000. Influence of measurement noise and electrode mislocalisation on EEG dipole-source localisation. Medical and Biological Engineering and Computing, 38(3), pp.287-296.
https://doi.org/10.1007/BF02347049

[79] Waldorp, L.J., Huizenga, H.M., Dolan, C.V. and Molenaar, P.C., 2001. Estimated generalized least squares electromagnetic source analysis based on a parametric noise covariance model [EEG/MEG]. IEEE Transactions on Biomedical Engineering, 48(6), pp.737-741.
https://doi.org/10.1109/10.923793

[80] Wan, J., Zhang, Z., Rao, B.D., Fang, S., Yan, J., Saykin, A.J. and Shen, L., 2014. Identifying the neuroanatomical basis of cognitive impairment in Alzheimer's disease by correlation-and nonlinearity-aware sparse Bayesian learning. IEEE transactions on medical imaging, 33(7), pp.1475-1487.
https://doi.org/10.1109/TMI.2014.2314712

[81] Wang, H., 2012. Bayesian graphical lasso models and efficient posterior computation. Bayesian Analysis, 7(4), pp.867-886.
https://doi.org/10.1214/12-BA729

[82] Wang, H., 2014. Coordinate descent algorithm for covariance graphical lasso. Statistics and Computing, 24(4), pp.521-529.
https://doi.org/10.1007/s11222-013-9385-5

[83] Wens, V., Marty, B., Mary, A., Bourguignon, M., Op De Beeck, M., Goldman, S., Van Bogaert, P., Peigneux, P. and De Tiège, X., 2015. A geometric correction scheme for spatial leakage effects in MEG/EEG seed‐based functional connectivity mapping. Human brain mapping, 36(11), pp.4604-4621.
https://doi.org/10.1002/hbm.22943

[84] Wills, A., Ninness, B. and Gibson, S., 2009. Maximum likelihood estimation of state space models from frequency domain data. IEEE Transactions on Automatic Control, 54(1), pp.19-33.
https://doi.org/10.1109/TAC.2008.2009485

[85] Wipf, D.P., Ramırez, R.R., Palmer, J.A., Makeig, S. and Rao, B.D., 2006. Automatic Relevance Determination for Source Localization with MEG and EEG Data. Technical Report, University of California, San Diego.

[86] Wipf, D.P. and Rao, B.D., 2007. An empirical Bayesian strategy for solving the simultaneous sparse approximation problem. IEEE Transactions on Signal Processing, 55(7), pp.3704-3716.
https://doi.org/10.1109/TSP.2007.894265

[87] Wipf, D. and Nagarajan, S., 2009. A unified Bayesian framework for MEG/EEG source imaging. NeuroImage, 44(3), pp.947-966.
https://doi.org/10.1016/j.neuroimage.2008.02.059

[88] Wipf, D.P., Owen, J.P., Attias, H.T., Sekihara, K. and Nagarajan, S.S., 2010. Robust Bayesian estimation of the location, orientation, and time course of multiple correlated neural sources using MEG. NeuroImage, 49(1), pp.641-655.
https://doi.org/10.1016/j.neuroimage.2009.06.083

[89] Wu, W., Nagarajan, S. and Chen, Z., 2016. Bayesian Machine Learning: EEG\/MEG signal processing measurements. IEEE Signal Processing Magazine, 33(1), pp.14-36.
https://doi.org/10.1109/MSP.2015.2481559

[90] Yuan, G., Tan, H. and Zheng, W.S., 2017. A Coordinate-wise Optimization Algorithm for Sparse Inverse Covariance Selection. arXiv preprint arXiv:1711.07038.
https://arxiv.org/abs/1711.07038

[91] Zhang, T. and Zou, H., 2014. Sparse precision matrix estimation via lasso penalized D-trace loss. Biometrika, 101(1), pp.103-120.
https://doi.org/10.1093/biomet/ast059

[92] Zhang, Z. and Rao, B.D., 2011. Sparse signal recovery with temporally correlated source vectors using sparse Bayesian learning. IEEE Journal of Selected Topics in Signal Processing, 5(5), pp.912-926.
https://doi.org/10.1109/JSTSP.2011.2159773

[93] Zhang, T. and Zou, H., 2014. Sparse precision matrix estimation via lasso penalized D-trace loss. Biometrika, 101(1), pp.103-120.

https://doi.org/10.1093/biomet/ast059

[94] Zhang, Y., Zhou, G., Jin, J., Zhao, Q., Wang, X. and Cichocki, A., 2016. Sparse Bayesian classification of EEG for brain–computer interface. IEEE transactions on neural networks and learning systems, 27(11), pp.2256-2267.
https://doi.org/10.1109/TNNLS.2015.2476656

[95] Zhang, Y., Wang, Y., Jin, J. and Wang, X., 2017. Sparse Bayesian learning for obtaining sparsity of EEG frequency bands based feature vectors in motor imagery classification. International journal of neural systems, 27(02), p.1650032.
https://doi.org/10.1142/S0129065716500325

# Appendix

## Mathematical notation

| | | |
|---|---|---|
| [A.1] | $x, \mathbf{X}, \mathbb{X}$ | The following symbols denote respectively a vector (bold italic lowercase) a matrix (bold capital) a set (double struck capital). |
| [A.2] | $x_m$ | Subscript indicating with lowercase script the $m$-th vector sample. |
| [A.3] | $X_{ij}, (X)_{ij}, x_i, (x)_i$ | Subscript indicating with lowercase the $ij$ ($i$) element of a matrix $\mathbf{X}$ (vector $x$). Light captions denote matrix (vectors) elements. |
| [A.4] | $N_p(x\|y, \mathbf{Z})$ | Normal distribution of a (p) size vector $x$ with mean $y$ and Covariance Matrix $\mathbf{Z}$. |
| [A.5] | $N_p^{\mathbb{C}}(x\|y, \mathbf{Z})$ | Circularly Symmetric Complex Normal distribution of a (p) size complex vector $x$ with complex mean $y$ and Complex Covariance Matrix $\mathbf{Z}$. |
| [A.6] | $exp(x\|y)$ | Exponential distribution of the scalar $x$ with parameter of shape $y$. |
| [A.7] | $Gamma(x\|y, z)$ | Gamma distribution of the scalar $x$ with parameters of shape $y$ and rate $z$. |
| [A.8] | $TGamma(x\|y, z, (a, b))$ | Truncated Gamma distribution of the scalar $x$ with parameters of shape $y$, rate $z$ and truncation interval $(a, b)$. |
| [A.9] | $\|\mathbf{X}\|$ | Determinant of a matrix $\mathbf{X}$. |
| [A.10] | $tr(\mathbf{X})$ | Trace of a matrix $\mathbf{X}$. |
| [A.11] | $\mathbf{X}^{-1}$ | Inverse of a matrix $\mathbf{X}$. |
| [A.12] | $\mathbf{X}^{\mathcal{T}}$ | Transpose of a matrix $\mathbf{X}$. |
| [A.13] | $\mathbf{X}^{\dagger}$ | Conjugate transpose of a matrix $\mathbf{X}$. |
| [A.14] | $\hat{\mathbf{X}}, \hat{x}$ | Estimator **Parameters** or **Hyperparameters** random matrix ($\mathbf{X}$) or vector ($x$). |
| [A.15] | $\check{\mathbf{X}}$ | Estimator of auxiliary magnitudes random matrix ($\mathbf{X}$), dependent on **Parameters** or **Hyperparameters** estimators. |
| [A.16] | $\hat{\mathbf{X}}^{(k)}, \check{\mathbf{X}}^{(k)}$ | Updates at the $k$-th iteration of estimators. |
| [A.17] | $\sum_{m=1}^{m}$ | Sum operator along index $m$. |

| [A.18] | $\prod_{m=1}^{m}$ | Product operator along index $m$. |
|---|---|---|
| [A.19] | $p(\mathbf{X})$ | Probability density function of a random variable **X**. |
| [A.20] | $p(\mathbf{X}|\mathbf{Y})$ | Conditional probability density function of a random variable **X** regarding the state of the variable **Y**. |
| [A.21] | $p(\mathbf{X},\mathbf{Y}|\mathbf{Z})$ | Conditional joint probability density function of random variables **X** and **Y** regarding the state of the variable **Z**. |
| [A.22] | $\mathbf{X}|\mathbf{Y} \backsim p(\mathbf{X}|\mathbf{Y})$ | Indicates that the variable **X** probability density function is conditioned to **Y**. |
| [A.23] | $\|\mathbf{X}\|_{i,\mathbf{A}_i}, i=1,2$ | L1 or L2 norm of the matrix **X** with weights or elementwise precisions defined by the mask matrix $\mathbf{A}_i$. |
| [A.24] | $\mathbf{I}_p, \mathbf{1}_p, \mathbf{0}_p$ | Denotes respectively Identity, Ones and Ceros matrices of size p. |
| [A.28] | $\odot, \oslash$ | Elementwise matrix product a division operators (Hadamard). |
| [A.29] | $argmin_{\mathbf{X}}\{f(\mathbf{X})\}$ or $argmax_{\mathbf{X}}\{f(\mathbf{X})\}$ | Extreme values of the scalar function $f$, correspondingly minimum or maximum, in the argument **X**. |
| [A.30] | $zeros_{\mathbf{X}}\{f(\mathbf{X})\}$ | Zeros of the scalar function $f$ in the argument **X**. |

Definition of variables

| [B.1] | $\mathbb{E}$ | Scalp Sensors (Electrodes) Space. |
|---|---|---|
| [B.2] | $\mathbb{M}$ | Random Samples space. |
| [B.3] | $\mathbb{G}$ | Discretized Gray Matter (Generators) Space. |
| [B.4] | p | Number of MEEG sensors at the scalp. |
| [B.5] | m | Number of data samples obtained from MEEG single frequency bin Fourier coefficients from a number (m) of segments. |
| [B.6] | q | Number of MEEG generators at the Cortex surface. |
| [B.7] | $\boldsymbol{v}_m$ | Complex size MEEG data Fourier coefficients sample vector for a single frequency component (observed variables or ***Data***). |
| [B.8] | $\boldsymbol{\iota}_m$ | Complex size MEEG source's Fourier coefficients sample vector for a single frequency component (unobserved variables or parameters). |
| [B.9] | **L** | Lead Field matrix of n × q size. |
| [B.10] | $\boldsymbol{\xi}_m$ | Complex Fourier coefficients vector for a single frequency component from MEEG forward model residuals (sensors' noise). |
| [B.11] | $\boldsymbol{\Sigma}_\iota$ | Complex size Hermitian and positive semidefinite matrix of EEG/MEG sources' Fourier coefficients (unobserved variables or ***Parameters***) Covariance matrix. |
| [B.12] | $\boldsymbol{\Theta}_\iota$ | Complex size Hermitian and positive semidefinite matrix of EEG/MEG source's Fourier coefficients |
| | | (unobserved variables or **Parameters**) Inverse Covariance matrix. |
| [B.13] | $\mathbf{\Sigma}_{\xi\xi}$ | Complex Hermitian and positive semidefinite matrix of EEG/MEG forward model residuals' Fourier coefficients (sensors' noise) Covariance matrix. |
| [B.14] | $\mathbf{R}$ | Known Complex Hermitian and positive semidefinite matrix of EEG/MEG sensors correlation structure. |
| [B.15] | $\sigma_\xi^2$ | Positive nuisance level hyperparameter $\sigma_e^2$. |
| [B.16] | $\Xi$ | General variable defining the set of hyperparameters. |
| [B.17] | $Q(\Xi, \hat{\Xi})$ | Data expected log likelihood, obtained after the expectation operation of the data and parameters log joint conditional probability density function over the parameters accounting for the parameters posterior density function with estimated values of the hyperparameters. |
| [B.18] | $\Pi(\mathbf{\Theta}_u, \mathbf{A})$ | Scalar general penalty function or exponent of the prior distribution Precision matrix $\mathbf{\Theta}_u$ parametrized in the regularization parameters or mask matrix $\mathbf{A}$. |
| [B.19] | $\lambda$ | Regularization parameters or tuning hyperparameters vector of the general penalty function. |
| [B.20] | $\breve{\mathbf{T}}^{(k)}$ | MEEG Data to Source activity Transference Operator. |
| [B.21] | $\breve{\mathbf{\Sigma}}_u^{(k)}$ | Complex Hermitian and positive semidefinite matrix of MEEG source Fourier coefficients (unobserved variables or parameters) posterior Covariance matrix. |
| [B.22] | $\breve{\mathbf{S}}_u^{(k)}$ | Complex Hermitian and positive semidefinite matrix of MEEG sources' Fourier coefficients (unobserved variables or parameters) empirical Covariance matrix. |
| [B.23] | $\breve{\mathbf{\Psi}}_u^{(k)}$ | Effective Sources Empirical Covariance (ESEC). It carries the information about sources correlations that will effectively influence the sources Covariance matrix estimator in the maximization step (sources Graphical Model solution), thus, it becomes the sources Covariance matrix estimator in the especial case of prior free model. |
| [B.24] | $\mathbf{S}_{vv}$ | Complex Hermitian matrix MEEG data Fourier coefficients Covariance matrix. |